\documentclass[sigconf]{acmart}

\AtBeginDocument{%
  }

\setcopyright{acmlicensed}
\copyrightyear{2018}
\acmYear{2018}
\acmDOI{XXXXXXX.XXXXXXX}

\acmConference[Conference acronym 'XX]{Make sure to enter the correct
  conference title from your rights confirmation emai}{June 03--05,
  2018}{Woodstock, NY}
\acmISBN{978-1-4503-XXXX-X/18/06}

\usepackage{graphicx}
\usepackage{caption}
\usepackage{subfigure} 
\usepackage{graphicx}
\usepackage{float}

\usepackage{amssymb}
\usepackage{amsmath}

\usepackage{algorithm} 
\usepackage{algorithmic} 
\usepackage{multirow} 
\usepackage{amsmath}
\usepackage{xcolor}
\usepackage{array}
\usepackage{multirow}
\usepackage{longtable}
\usepackage{rotating}
\usepackage{booktabs} 
\usepackage{bigstrut}

\usepackage{enumitem}
\usepackage{xspace}
\usepackage{balance}
\usepackage{hyperref}

\newcommand\mymodel{{CLP}\xspace}
\newcommand{\clp}{{CLP}}

\newlength{\myHeight}
\begin{document}

\title{How to Bridge Spatial and Temporal Heterogeneity in Link Prediction? A Contrastive Method}

\author{Yu Tai}
\orcid{0000-0002-6920-9801}
\affiliation{%
	\institution{School of Cyberspace Science, \\Harbin Institute of Technology}
	\streetaddress{No.92 Xidazhi Street}
	\city{Harbin}
	\country{China}
	\postcode{150001}
}
\email{taiyu@hit.edu.cn}

\author{Xinglong~Wu}
\orcid{0000-0003-4139-052X}
\affiliation{%
	\institution{School of Cyberspace Science, \\Harbin Institute of Technology}
	\streetaddress{No.92 Xidazhi Street}
	\city{Harbin}
	\country{China}
	\postcode{150001}
}
\email{xlwu@stu.hit.edu.cn}

\author{Hongwei~Yang}
\orcid{0000-0002-8386-0131}
\affiliation{%
	\institution{School of Cyberspace Science, \\Harbin Institute of Technology}
	\streetaddress{No.92 Xidazhi Street}
	\city{Harbin}
	\country{China}
	\postcode{150001}
}
\email{yanghongwei@hit.edu.cn}

\author{Hui~He}
\orcid{0000-0002-6494-775}
\authornote{Corresponding author}
\affiliation{%
	\institution{School of Cyberspace Science, \\Harbin Institute of Technology}
	\streetaddress{No.92 Xidazhi Street}
	\city{Harbin}
	\country{China}
	\postcode{150001}
}
\email{hehui@hit.edu.cn}

\author{Duanjing~Chen}
\orcid{0009-0009-1432-3111}
\affiliation{%
	\institution{School of Cyberspace Science, \\Harbin Institute of Technology}
	\streetaddress{No.92 Xidazhi Street}
	\city{Harbin}
	\country{China}
	\postcode{150001}
}
\email{duanjingchen@stu.hit.edu.cn}

\author{Yuanming~Shao}
\orcid{0009-0009-3020-3494}
\affiliation{%
	\institution{School of Cyberspace Science, \\Harbin Institute of Technology}
	\streetaddress{No.92 Xidazhi Street}
	\city{Harbin}
	\country{China}
	\postcode{150001}
}
\email{ymshao@stu.hit.edu.cn}

\author{Weizhe~Zhang}
\authornotemark[1]
\orcid{0000-0003-4783-876X}
\affiliation{%
	\institution{Computer Science and Technology, \\Harbin Institute of Technology at Shenzhen}
	\city{Shenzhen}
	\country{China}
	\postcode{518067}
}
\affiliation{%
	\institution{Department of New Networks,
		\\Peng Cheng Laboratory}
	\city{Shenzhen}
	\country{China}
	\postcode{518055}
}
\email{wzzhang@hit.edu.cn}

\renewcommand{\shortauthors}{Yu Tai, Xinglong Wu, and et al.}

\begin{abstract}
  Temporal Heterogeneous Networks play a crucial role in capturing the dynamics and heterogeneity inherent in various real-world complex systems, rendering them a noteworthy research avenue for link prediction.
  However, existing methods fail to capture the fine-grained differential distribution patterns and temporal dynamic characteristics, which we refer to as spatial heterogeneity and temporal heterogeneity.
  To overcome such limitations, we propose a novel \textbf{C}ontrastive Learning-based \textbf{L}ink \textbf{P}rediction model, \textbf{CLP}, which employs a multi-view hierarchical self-supervised architecture to encode spatial and temporal heterogeneity.
  Specifically, aiming at spatial heterogeneity, we develop a spatial feature modeling layer to capture the fine-grained topological distribution patterns from node- and edge-level representations, respectively.
  Furthermore, aiming at temporal heterogeneity, we devise a temporal information modeling layer to perceive the evolutionary dependencies of dynamic graph topologies from time-level representations.
  Finally, we encode the spatial and temporal distribution heterogeneity from a contrastive learning perspective, enabling a comprehensive self-supervised hierarchical relation modeling for the link prediction task.
  Extensive experiments conducted on four real-world dynamic heterogeneous network datasets verify that our \mymodel consistently outperforms the state-of-the-art models, demonstrating an average improvement of 10.10\%, 13.44\% in terms of AUC and AP, respectively.
\end{abstract}

\ccsdesc[500]{Networks~Network dynamics}

\keywords{Link prediction, Temporal heterogeneous graph, Graph representation learning, Contrastive learning}

\received{20 February 2007}
\received[revised]{12 March 2009}
\received[accepted]{5 June 2009}

\maketitle

\section{Introduction}
Contemporary information networks such as social networks~\cite{WangLH24} and biological systems~\cite{chen2022graph} are becoming increasingly complex.
These networks often comprise multi-typed nodes and connections, undergo continuous temporal evolution, making the link prediction in such complex networks a long-standing challenge. 
Specifically, the link prediction task aims to predict the likelihood of future connections between arbitrary nodes~\cite{WangLSCFLG24,YaoL24,WuSGG22}, which captures the evolution of heterogeneous networks and stores the temporal details of the node embeddings, simulating intricate and expressive semantics for real-world systems, including Social Recommendations~\cite{wu2023pda}, Traffic Management~\cite{JinRZJGLMW22}, Medical Health~\cite{wang2021multi} and Network Biology~\cite{GuzziZ22}.
Aiming at modeling the dynamics and complex relationships between entities, link prediction models are primarily designed to portray the topological relationship between heterogeneous snapshots and the evolving progress along chronological order, revealing the distribution patterns within complex Temporal Heterogeneous Networks (THNs).

The primary challenges in link prediction tasks revolve around heterogeneous entity relationship modeling and dynamic snapshot variability modeling. Current link prediction methods in literature typically segment dynamic snapshot sequences chronologically and address the complex entity relationships existing in each snapshot.
We term such challenges as \textit{spatial complexity} and \textit{temporal complexity}.
(1) \textit{Spatial complexity}~\cite{wang2019heterogeneous,zhang2023page} highlights the complex heterogeneous static relationships between multi-typed entities in complex networks, primarily modeling the diverse co-occurrence paradigm through heterogeneous network embedding approaches.
Specifically, Meta-Path-based approaches~\cite{qiu2018network, dong2017metapath2vec, fu2017hin2vec,zhao2023link} construct meta-paths within individual snapshots to excavate the heterogeneous information.
On the other hand, Attribute-based methods~\cite{hu2020heterogeneous,li2020type,zhang2023dynamic1} focus on incorporating multiple rich attributes~\cite{xu2021topic,peng2023th} and merging neighbor attributes~\cite{hu2020heterogeneous,li2020type} to enhance the node embedding process.
(2) \textit{Temporal complexity}~\cite{pareja2020evolvegcn,pham2021comgcn,fan2022heterogeneous} mainly exploits the dynamic distribution changes in snapshot sequences. Temporal approaches focus on tracking the continuous evolution across chronological snapshots, primarily classified into sequential and graph methods.
Sequential methods~\cite{hao2020dynamic,jiang2023dnformer} learn from time-ordered snapshot sequences, capturing evolutionary dependencies between different snapshots based on Recurrent Neural Networks (RNNs)~\cite{chen2019lstm} and attention mechanism~\cite{zhang2023attentional}.
Graph methods~\cite{lei2019gcn,yang2019advanced,XuRKKA20,yang2022few} aggregate embeddings of dynamic nodes, encoding the appearing or disappearing network features continuously over time through Graph Neural Networks (GNNs).

\begin{figure}[!htbp]
	\centering
	\includegraphics[width=\linewidth]{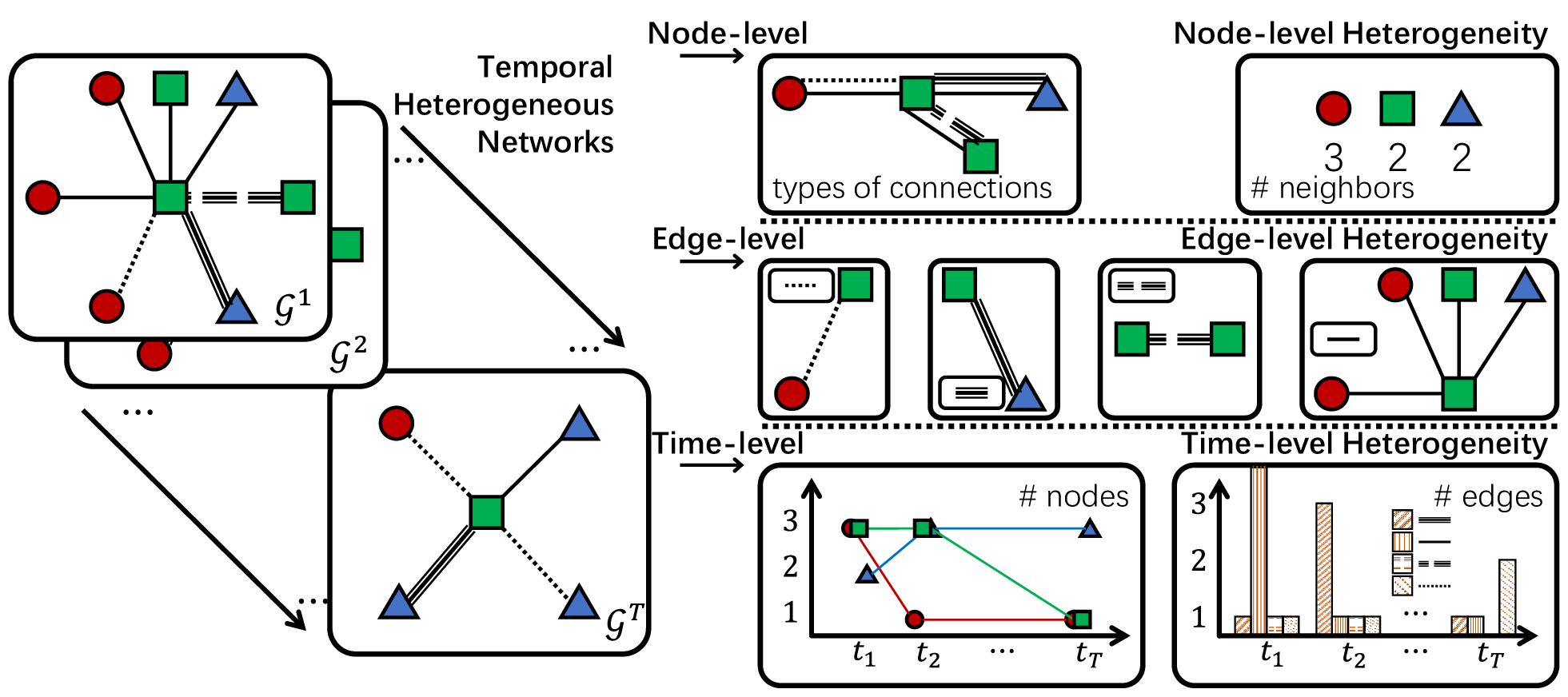}
	\caption{An Illustrative Example.}
	\label{fig:illustration}
\end{figure}

Despite the effectiveness, a prominent drawback of these methods is that they model dynamic heterogeneous representations in a coarse-grained manner and only focus on the representation paradigm, \textit{but ignore the universally-distributed differential relations in THNs}, thus resulting in the suboptimal performance in link prediction tasks.
We reckon that by leveraging such differential relations and bridging temporal and spatial heterogeneity, it becomes feasible to portray the comprehensive and detailed dynamic and diversified characteristics, thereby enhancing link prediction performance.
Specifically, focusing on the aforementioned \textit{temporal} and \textit{spatial complexity}, the fine-grained differential relations between different nodes and edges (\textit{spatial}) and the evolution paradigm distinctions (\textit{temporal}) play a crucial role in representation learning and significantly influence link prediction performance.

We illustrate such fine-grained differential distributions at the node-, edge-, and time-level in Figure~\ref{fig:illustration}.
From the node-level propagation, taking the static THN snapshot $\mathcal{G}^1$ as an example, different propagation and aggregation paradigms convey differential information.
If we take the quadrangle (\includegraphics[height=\myHeight]{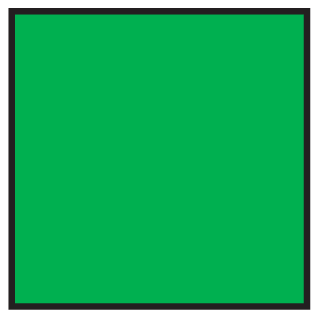}) as the ego node, from the perspective of connection types, the propagation priority sequence for \includegraphics[height=\myHeight]{quadrangle.eps} reveals:
\includegraphics[height=\myHeight]{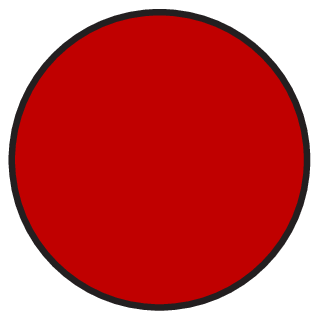} $=$
\includegraphics[height=\myHeight]{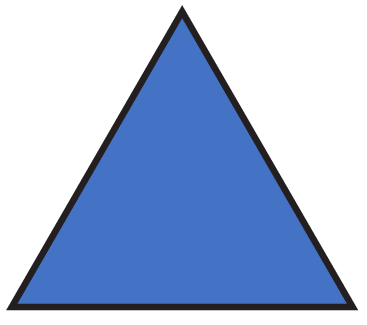} $=$
\includegraphics[height=\myHeight]{quadrangle.eps}.
Meanwhile, from the perspective of single-type neighbor numbers, the priority manifests:
\includegraphics[height=\myHeight]{circle.eps} $>$
\includegraphics[height=\myHeight]{quadrangle.eps} $=$
\includegraphics[height=\myHeight]{triangle.eps},
reflecting significant distribution heterogeneity among different propagation patterns.
Analogously, targeting edge-level propagation, different edge types form the holistic propagation paradigm, yet convey unique propagation information.
However, the heterogeneity between different edge-level propagation lacks sufficient attention.
Finally, from time-level propagation, significant variations occur due to diverse inspection degrees (e.g., number of nodes, variations of edges), highlighting the transformative information conveyed by different time-level propagation patterns.
We term such differential distribution in THNs as `\textit{spatial heterogeneity}' and `\textit{temporal heterogeneity}', which are crucial factors for link prediction modeling but have been rarely addressed in related research.

As is illustrated in the above example, different entities in complex networks possess variously-grained distribution differentiation, existing among nodes, edges and varying along chronological order.
However, conventional methods ignore the modeling of such heterogeneity differentiation, which introduces the first challenge in this work, i.e., 
\textbf{CH1:} \textit{How to capture the heterogeneity differentiation existing in link relation networks?}
To represent such distribution heterogeneity in link prediction, it is essential to characterize the fine-grained intrinsic topological distribution in the link graph.
Accordingly, we resort to Self-supervised Learning~\cite{chen2020simple} to investigate the inherent inter-relation in temporal heterogeneous graphs.

In addition, to bridge the discrepancy elimination and topological exploration module in link prediction, we need to resolve the second challenge, i.e.,
\textbf{CH2:} \textit{How to integrate different granularity of distribution discrepancy?}
To address this challenge, we first design a heterogeneous temporal graph to absorb both structural distribution patterns and sequential evolutionary paradigms.
Subsequently, we propose a contrastive hierarchical heterogeneity differentiation module to absorb the intrinsic inter-relation from node-, edge-, and time-level, respectively.
Leveraging such hierarchical contrastive module, we implement a fine-grained multi-view entity relation extraction functionality.

To summarize, our main contributions are as follows:
\begin{itemize}[leftmargin=*]
	\item Targeting CH1, we propose a three-layer hierarchical contrastive entity relation extraction module to enable multi-view discrepancy elimination functionality, thereby bridging the spatial and temporal heterogeneity in link prediction scenarios.
	\item Targeting CH2, we design a heterogeneous temporal graph network to absorb sequential and structural distribution paradigms and comprehensively eliminate discrepancies from various perspectives.
	Specifically, we depict the structural distribution differentiation paradigms with node- and edge-level graph networks and propose a dual-channel sequential module to capture different sequential reliance among snapshots.
	\item We conduct extensive experiments on four benchmark datasets to predict temporal links between two entities. The experimental results indicate that \mymodel achieves superior prediction performance compared to the existing state-of-the-art link prediction methods.
\end{itemize}


\section{Preliminaries}
\label{Sec:Preliminaries}

\subsection{Problem Formalization}
\noindent
\textbf{Definition 1 (Heterogeneous Network (HN)).} Let $\mathcal{G}(V, E, A_v, A_e)$ be an undirected graph, where $V=\{v_1, v_2, \cdots, v_N\}$, $E=\{e_1, e_2, \cdots,\\ e_M\}$, $A_v$, and $A_e$ denote the set of nodes, edges, node types and edge types, respectively. Each node $v_i \in V$ and edge  $e_{j} \in E$ are associated with their corresponding node type $\varphi(v_i) \in A_v$ and edge type $\varphi(e_{j}) \in A_e$, and $|A_v|+|A_e|>2$.

\noindent
\textbf{Definition 2 (Temporal Heterogeneous Network (THN)).}
Let $\mathcal{G}(V, E, T, \mathbf{X})$ denote an undirected heterogeneous graph, comprising a sequence of heterogeneous network snapshots at multiple timesteps, i.e., $\mathcal{G}=\{\mathcal{G}^1, \mathcal{G}^2, \cdots, \mathcal{G}^T\}$, where $\mathcal{G}^t(V^t, E^t)$ is the network snapshot graph at time step $t$.
Here, $V^t=\left\{v_1^t, v_2^t, \cdots, v_{|V^t|}^t\right\} \subseteq V$ and $E^t=\left\{e_1^t, e_2^t, \cdots, e_{|E^t|}^t\right\} \subseteq E$ denote the node set and edge set at moment $t$, respectively. $T$ represents the total number of snapshots, and $V = \bigcup_{t=1}^{T}V^{t}$, $E = \bigcup_{t=1}^{T}E^{t}$.
For any node $a \in V$, a fixed-size feature vector $\mathbf{x}_{a} \in \mathbb{R}^{d}$ is given for node representation, and $\mathbf{X} = \left\{\mathbf{x}_{a}\right\}_{a \in V}$ denote the feature matrix for all nodes.

\noindent
\textbf{Temporal Heterogeneous Network Link Prediction Formalization.}
Our model aims to predict the possible link between two target nodes.
To address this, we formulate the temporal heterogeneous network link prediction problem as follows:
given a sequence of temporal heterogeneous graphs $\mathcal{G} = \left\{\mathcal{G}^1, \mathcal{G}^2, \cdots, \mathcal{G}^{T}\right\}$ and the target link $e = (a, b)$, where $a, b  \in V$,
our objective is to determine the likelihood of $e$ existing in $\mathcal{G}^{T+1}$ at  time step $T+1$ by assessing the similarity between representations of node $a$, and $b$ at time $T$ (denoted as $\mathbf{u}_a ^ T$ and $\mathbf{u}_b^T$).

The key mathematical symbols and definitions relevant to this article are summarized in Table \ref{tab:symbols}.

\begin{table}[!t]
	\scriptsize
	\centering
	\caption{Key Mathematical Notations.}
	\resizebox{\linewidth}{!}{
		\begin{tabular}{cc}
			\toprule
			Symbol               & Description \\
			\midrule
			$\mathcal{G}=\{\mathcal{G}^1, \mathcal{G}^2, \cdots, \mathcal{G}^T\}$ & Heterogeneous network sequences at different moments.\\
			$\mathcal{G}^t$      & The heterogeneous snapshot graph at $t$. \\
			${V}^t$              & The set of nodes in $\mathcal{G}^t$. \\
			${E}^t$              & The set of edges in $\mathcal{G}^t$. \\
			$\mathcal{G}^{rt}$   & The sub-network at time $t$ and with edge type $r$. \\
			$T$                  & The maximum number of graph snapshots. \\
			$\beta^{rt}_{a,b}$   & The attention score between nodes $a$ and $b$ in the type $r$ subgraph at time $t$.\\
			$\alpha^{rt}_{a,b}$  & The attention weight between nodes $a$ and $b$ in the type $r$ subgraph at time $t$. \\
			$\mathbf{u}^{rt}_a$  & The Node-level representation of node $a$ in the type $r$ subgraph at time $t$. \\
			$\gamma_{a}^{r t}$   & The attention score of node $a$ for edge type $r$ in the $t$-th snapshot.\\
			$\delta_{a}^{r t}$   & The attention weight of node $a$ for edge type $r$ in the $t$-th snapshot. \\
			$\mathbf{u}^t_a$     & The Edge-level representation of node $a$ at time $t$. \\
			$\mathbf{u}_a^L$     & The long-term temporal representation of node $a$. \\
			$\mathbf{u}_a^S$     & The short-term temporal representation of node $a$. \\
			$h$                  & The number of heads in the attention network. \\
			\bottomrule
		\end{tabular}
	}
	\label{tab:symbols}
\end{table}

\section{Methodology}
\label{Sec:Method}

In this section, we elaborate on the detailed architecture of our \mymodel to learn and encode the spatial and temporal heterogeneity in link predictions, introducing the hierarchical contrastive relation extraction modules from node-, edge-, and time-level, respectively.

The primary objective of our \mymodel is to learn a deep representation of a THN from various dynamic node and edge types to express the spatial discrepancy and temporal nonuniformity. To this end, we design \mymodel with a hierarchical architecture to capture the distribution heterogeneity, including 
(1) \textit{Spatial Feature Modeling Layer},
(2) \textit{Temporal Information Modeling Layer}, and
(3) \textit{Output Layer}.

\textbf{(1) Spatial Feature Modeling Layer:} First, we propose a two-layer hierarchical Graph Attention Network (GAT) to represent diverse types of edges and nodes within the THN from both node- and edge-level perspectives. Additionally, we introduce a contrastive representation method to differentiate feature heterogeneity at the node and edge levels, enhancing our ability to capture spatial heterogeneity.

\textbf{(2) Temporal Information Modeling Layer:}  Then, we deploy Long Short-Term Memory (LSTM) and Gated Recurrent Units (GRU) models to independently analyze temporal snapshot pattern, capturing the long-term and short-term dependencies between snapshots, respectively. Additionally, we implement contrastive learning strategies to bridge differences between these two sequence learning paradigms, thus preserving the temporal heterogeneity.

\textbf{(3) Output Layer:} Finally, we calculate the similarity between node $a$ and node $b$ to 
represent the target link $e = (a,b)$. This measurement is then incorporated into a comprehensive loss function to estimate the probability of the existence of the target link.

\begin{figure*}[!t]
	\small
	\centering
	\includegraphics[width=.8\linewidth]{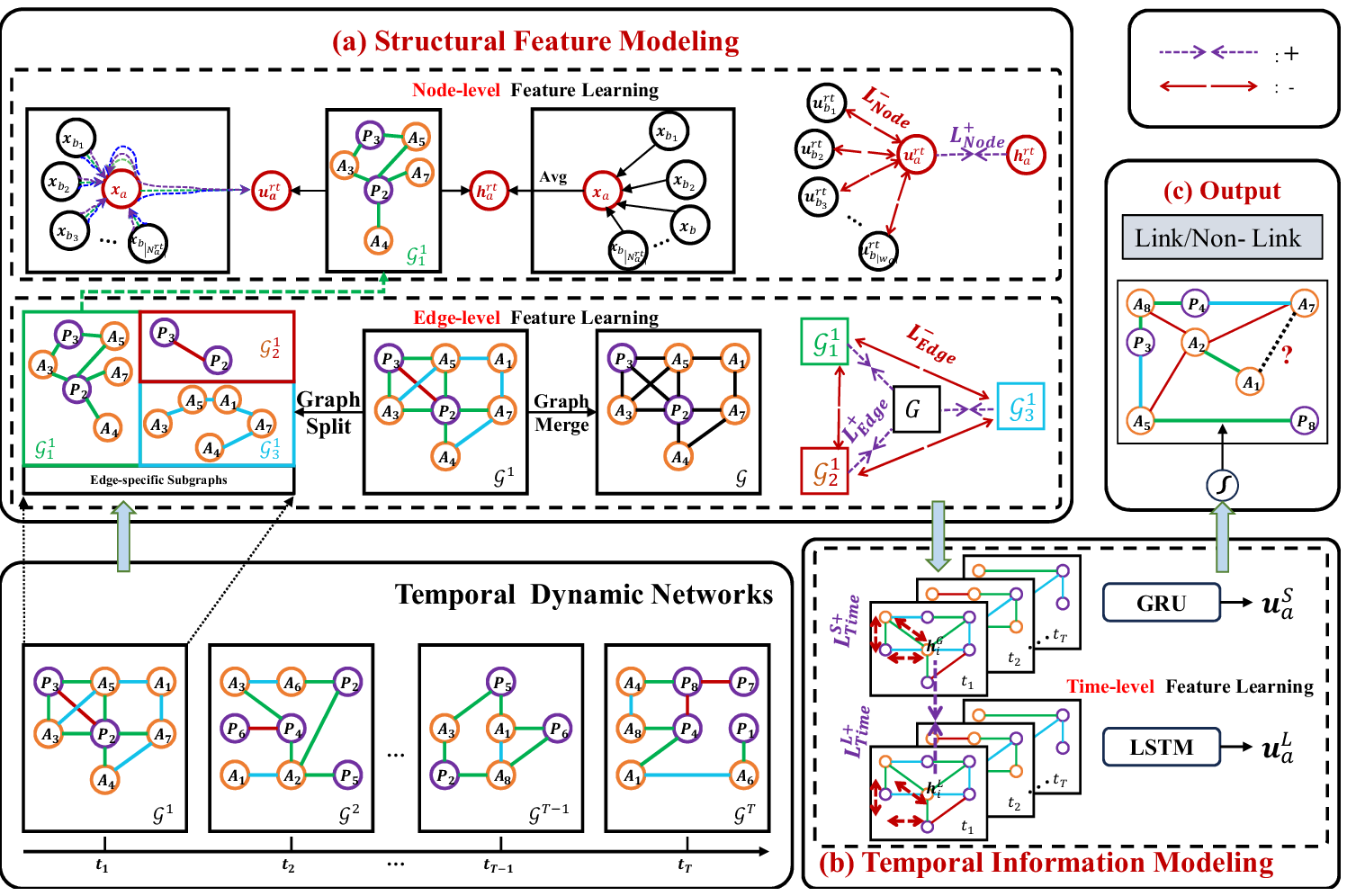}
	\caption{The Architecture of Our \mymodel Model.}
	\label{Fig:structure}
\end{figure*}

The aforementioned layers of our proposed \mymodel are shown in Figure~\ref{Fig:structure} with elaborate interpretations provided in the following subsections.

\subsection{Spatial Feature Modeling Layer}
\label{GNN}

The Spatial Feature Modeling Layer aims to capture the structural distribution patterns and eliminate the discrepancy between nodes and edges in each static snapshot $\mathcal{G}^{t} \in \mathcal{G}$.
Specifically, we initially partition the static snapshots $\mathcal{G}=\{\mathcal{G}^1, \mathcal{G}^2, \cdots, \mathcal{G}^T\}$ into type-specific sub-networks based on different edge types.
Then, we devise a two-layer hierarchical GAT to represent diverse types of nodes and edges within the THN from the perspectives of both node- and edge-levels in the Node-level Feature Learning Module and Edge-level Feature Learning Module, respectively.
Additionally, we devise the contrastive representation heterogeneity differentiation modeling in the node- and edge- feature learning module to model the \textit{spatial heterogeneity}.

\subsubsection{Node-level Feature Learning Module}
For each graph snapshot $\mathcal{G}^t \in \mathcal{G}$, we partition it into several subgraphs based on the edge type $r \in R$. The attention score $\beta^{rt}_{ab}$ between node $a$ and $b$ in the $r-$th type subgraph of the $t-$th static snapshot $\mathcal{G}^t$ is expressed as follows:
\begin{equation}
	\beta^{rt}_{a,b} = \left(\sigma\left({\mathbf{A}^{rt}}^{\top} \left[\mathbf{W}^{rt}  \mathbf{x}_{a} \| \mathbf{W}^{rt}  \mathbf{x}_{b}\right]\right)\right),
	\label{N-S}
\end{equation}
where $\mathbf{x}_a$ and $\mathbf{x}_b$ are used to initialize node $a$ and $b$;
$\mathbf{A}^{rt}$ and $\mathbf{W}^{rt}$ represent the learnable attention weight vector and mapping matrix, specific to $r$ type subgraphs of $\mathcal{G}^t$;
The function $\sigma\left( \cdot \right)$ denotes the activation function and  $\|$ signifies the concatenation operation.
The formalized attention weight parameter $\alpha^{rt}_{ij}$ between node $a$ and $b$ in the $r$ type subgraph of $\mathcal{G}^t$ is defined as follows:
\begin{equation}
	\alpha^{rt}_{a,b} = \frac{\exp (\beta^{rt}_{a,b})}{\sum_{k \in \mathcal{N}^{rt}_a} \exp (\beta^{rt}_{a,k})},
	\label{N-W}
\end{equation}
where $\mathcal{N}^{rt}_a$ represents the neighbors of node $a$ falling into the $r-$th type in $\mathcal{G}^t$.
The representation $\mathbf{u}_{a}^{r t}$ is obtained for each node by the weighted summation of the neighbor nodes, which attentively propagates and aggregates the node embeddings as follows:

\begin{equation}
	\begin{aligned}
		\mathbf{u}_{a}^{r t}=\sigma\left(\sum_{b \in \mathcal{N}_{a}^{r t}} \alpha_{a,b}^{r t}  \mathbf{W}^{rt} \mathbf{x}_{b}\right).
	\end{aligned}
	\label{N-H1}
\end{equation}

The GAT model highlights the diverse node representations but restricts the expression of a node's intrinsic embedding. Average embedding pooling effectively preserves the unique characteristics of node representations. Consequently, we implement a linear aggregation operation using GNN to represent nodes in the unified graph as follows:

\begin{equation}
	\mathbf{h}_a^{r t} = \mathbf{W}^{r t} \mathbf{x}_{a}+\frac{1}{\sqrt{\left| \mathcal{N}_a^{rt} \right| \cdot \left| \mathcal{N}_b^{rt} \right|}} \sum_{b \in \mathcal{N}_a^{rt}} \mathbf{W}^{rt} \mathbf{x}_{b}.
	\label{N-H2}
\end{equation}

\noindent
\textbf{Node-level Heterogeneity Differentiation Modeling.}
To enhance the representation of each node, we employ a node-level contrastive learning approach to ensure that augmented representations of the same node are similar (intra-similarity) and simultaneously differ from representations of other nodes (inter-dissimilarity).

Additionally, we implement a node-level InfoNCE loss function~\cite{abs-1807-03748}. This function guarantees that representations $\mathbf{u}_a^{rt}$ and $\mathbf{h}_{a}^{rt}$, derived from GAT and GNN for the same node $a$, are similar. Conversely, representation $\mathbf{u}_b^{rt}$, derived from GAT for a different node $b$, remains distinct. Within this framework, the pair ($\mathbf{u}_{a}^{rt}$, $\mathbf{h}_{a}^{rt}$) constitutes a positive sample pair, while the pair ($\mathbf{u}_a^{rt}$, $\mathbf{u}_b^{rt}$) functions as a negative sample pair, as specified in Eq.~\eqref{loss-N}:


\begin{equation}
		\resizebox{.9\linewidth}{!}{$
	\begin{aligned}
		\mathcal{L}^{{+}}_{{Node }}=-\sum_{t=1}^{T} \sum_{r=1}^{R} \sum_{a=1}^{\left|V^{t}\right|} \log \frac{\exp \left(\operatorname{sim}\left(\mathbf{u}_{a}^{r t}, \mathbf{h}_{a}^{r t}\right) / \tau\right)}{\sum_{b \in \mathcal{N}_{a}^{r t}} \exp \left(\operatorname{sim} \left(\mathbf{u}_{a}^{r t}, \mathbf{h}_{b}^{r t}\right) / \tau\right)},\\
	\mathcal{L}^{{- }}_{{Node }}=-\sum_{t=1}^T \sum_{r=1}^R \sum_{a=1}^{\left|V^{t}\right|} \sum_{\substack{b \in \mathcal{N}_a^{rt} \land b \neq a}} \log \frac{\exp \left(\operatorname{sim} \left(\mathbf{u}_a^{r t}, \mathbf{u}_b^{r t}\right) / \tau\right)}{\sum_{k \in \mathcal{N}_a^{r t}} \exp \left(\operatorname{sim} \left(\mathbf{u}_a^{r t}, \mathbf{u}_k^{r t}\right) / \tau\right)},
	\end{aligned}
$}
	\label{loss-N}
\end{equation}
where $\tau$ represents the temperature hyperparameter, and the dot product operation is used to compute the similarity between two vectors through $\operatorname{sim} \left( \cdot, \cdot \right)$, i.e., $\operatorname{sim} \left( \mathbf{u}_a^{r t}, \mathbf{u}_b^{r t} \right) = {\mathbf{u}_a^{rt}}^{\top} \cdot \mathbf{u}_b^{r t} $. Additionally, $\left|V^{t}\right|$, $T$, and $R$ are the size of $\mathcal{G}^t$, the size of snapshots and the number of edge types, respectively.

\subsubsection{Edge-level Feature Learning Module}
The Node-level Feature Learning module captures information specific to one edge type. However, heterogeneous networks typically comprise multiple edge types. To equip the information from all edge types at each node, we devise the Edge-level Feature Learning module, which determines the importance weights for different edge types. This module aggregates different forms of information for a specific type, thereby generating the node's embedding enriched with heterogeneous edge information. Specifically, each node's embedding vector undergoes a nonlinear mapping. The attention weight $\delta_{a}^{r t}$ between each edge type $r$ and node $a$ in the $t$-th snapshot graph is obtained through the softmax activation function, as described in Eq.\eqref{edge-s}:
\begin{equation}
	\begin{aligned}
		\gamma_{a}^{rt} &= \mathbf{z}^{\top} \cdot \sigma\left(\mathbf{W}^t \mathbf{u}^{r t}_a + \mathbf{b}\right), \\
		\delta_{a}^{r t} &= {\exp \left(\gamma_{a}^{r t}\right)} / {\sum\nolimits_{r^{\prime} \in R} \exp \left(\gamma_{a}^{r^{\prime} t}\right)},
	\end{aligned}
	\label{edge-s}
\end{equation}
where $\mathbf{z}$, $\mathbf{W}^t$, and $\mathbf{b}$ represent the trainable attention weight vector, weight matrix, and bias vector, respectively. $\sigma$ denotes a non-liner activation function. Then, the representation of each node $\mathbf{u}_{a}^{ t}$ in $\mathcal{G}^t$ is derived by aggregating edge-specific information through the weighted summation:

\begin{equation}
	\mathbf{u}_a^t=\sum_{r=1}^R \delta_a^{r t} \mathbf{u}_a^{r t}.
	\label{E-H1}
\end{equation}

Analogously, the GAT model highlights the diverse representation of various edge types; however, it limits the expression of the edge type itself. Average embedding pooling effectively preserves the intrinsic characteristics of edge type representations. Consequently, a linear aggregation operation is implemented through GNN as follows:
\begin{equation}
	\mathbf{h}_{a}^{t}=
	\frac{1}{R} \sum_{r=1}^{R}
(
	\mathbf{u}_{a}^{rt}+
	\frac{1}
	{\sqrt{|\mathcal{N}_{a}^{r t}| \cdot |\mathcal{N}_{b}^{r t}|}
	}
	\sum_{b \in \mathcal{N}_{a}^{rt}} \mathbf{u}_{b}^{r t}
),
	\label{E-H2}
\end{equation}
where $\mathcal{N}_{a}^{rt}$ denotes the neighbors of node $a$ with $r$ edge type in $\mathcal{G}^t$.

\noindent
\textbf{Edge-level Heterogeneity Differentiation Modeling.} To refine the representation of each edge-specific node, we employ an edge-level contrastive learning approach to ensure that augmented representations of identical subgraphs demonstrate intra-similarity, while those of different subgraphs exhibit inter-dissimilarity.

Furthermore, to confirm that the edge-specific embedding $\mathbf{u}_a^t$ for node $a$ is similar to its embedding $\mathbf{h}_a^t$ in the unified graph and dissimilar to the aggregated representation $\mathbf{u}_b^t$ for node $b$, we define the edge-level InfoNCE loss function considering the heterogeneous and unified aggregated representations of the same node as a positive sample pair ($\mathbf{u}_{a}^{t}$, $\mathbf{h}_{a}^{t}$), while forming a negative sample pair ($\mathbf{u}_{a}^{t}$, $\mathbf{u}_{b}^{t}$) with heterogeneous aggregated representations from different nodes, as expressed in Eq.~\eqref{loss-E}:
\begin{equation}
	\resizebox{.9\linewidth}{!}{$
		\begin{aligned}
			\mathcal{L}^{{+}}_{ {Edge }} &= -\sum_{t=1}^{T} \sum_{a=1}^{\left|V^{t}\right|} \log \frac{\exp \left(\operatorname{sim}\left(\mathbf{u}_{a}^{t}, \mathbf{h}_{a}^{t}\right) / \tau\right)}{\sum_{b \in \mathcal{N}_a^t} \exp \left( \operatorname{sim} \left(\mathbf{u}_{a}^{t}, \mathbf{h}_{b}^{t}\right) / \tau\right)} , \\
			\mathcal{L}^{ {-}}_{ {Edge }} &= -\sum_{t=1}^{T} \sum_{a=1}^{\left|V^{t}\right|} \sum_{\substack{b \in \mathcal{N}_a^t \land b \neq a}}  \log \frac{\exp \left(\operatorname{sim} \left(\mathbf{u}_{a}^{t}, \mathbf{u}_{b}^{t}\right) / \tau\right)}{\sum_{k \in \mathcal{N}_{a}^{t}} \exp \left(\operatorname{sim} \left(\mathbf{u}_{a}^{t}, \mathbf{u}_{k}^{t}\right) / \tau\right)},
		\end{aligned}
		$}
	\label{loss-E}
\end{equation}
where $\mathcal{N}_{a}^{t} = \bigcup_{r} {\mathcal{N}_{a}^{rt}}$ represents the neighbors of node $a$ in $\mathcal{G}^t$ including all edge types.
By alienating the neighbors in the unified graph $\mathcal{G}^{t}$, we not only strengthen contrastive relationships in Eq. \eqref{loss-N} for intra-relation neighbors, but also alienate the inter-relation neighbors with different edge types.

\subsection{Temporal Information Modeling Layer}
\label{Temporal}
The Temporal Information Modeling Layer aims to address the variability of temporal sequence information across different sequence modeling contexts, where various modeling techniques can capture distinct sequence patterns. Specifically, we employ a dual-channel architecture to learn different sequential dependency paradigms\textemdash (1) in the long-term channel, we deploy LSTM to explore the inherent inter-dependencies in long-term temporal evolution and (2) in the short-term channel, we apply GRU to analyze interactions between adjacent snapshots in short-term evolution. However, previous studies have overlooked the heterogeneity between these long-term and short-term dependencies. To this end, we propose a contrastive learning approach to emphasize the differences between diverse sequence learning paradigms, thereby highlighting temporal heterogeneity. We deploy LSTM and GRU to represent the temporal patterns for learning long and short dependencies, expressed as $\mathbf{u}_a^L$ and $\mathbf{u}_a^S$, respectively:
\begin{equation}
	\mathbf{u}_a^L \leftarrow f_{\textit{LSTM}}\left(\left\{\mathbf{u}_a^t \right\}_{t=1}^T\right),
	\label{T-L}
\end{equation}
\begin{equation}
	\mathbf{u}_a^S \leftarrow f_{\textit{GRU}}\left(\left\{ \mathbf{u}_a^t \right\}_{t=1}^T \right),
	\label{T-G}
\end{equation}
where $\left\{\mathbf{u}_a^t\right\}_{t=1}^T= \{ \mathbf{u}_a^1, \mathbf{u}_a^2, \dots, \mathbf{u}_a^T  \}$ denotes the node $a$'s embeddings across various snapshots of all time points.
Then, the embeddings for node $a$ throughout all snapshots are derived. 

\noindent
\textbf{Time-level Heterogeneity Differentiation Modeling.}
To bridge the nonuniformity of temporal sequences between long- and short-term sequence learning spaces, we propose a time-level contrastive learning method to approach the latent long-term and short-term latent sequential representations, respectively:

%

\begin{equation}
		\begin{aligned}
	\mathcal{L}^{L+}_{Time }=\sum_{a=1}^{\left|V^T\right|} \log \frac{\exp \left(\operatorname{sim} \left(\mathbf{u}_a^L, \mathbf{u}_a^S\right) / \tau\right)}{\sum_{b \neq a} \exp \left(\operatorname{sim} \left(\mathbf{u}_a^L, \mathbf{u}_b^L\right) / \tau\right)},\\
		\mathcal{L}^{S+}_{Time }=\sum_{a=1}^{\left|V^T\right|} \log  \frac{\exp \left(\operatorname{sim} \left(\mathbf{u}_a^S \cdot \mathbf{u}_a^L\right) / \tau\right)}{\sum_{b \neq a} \exp \left(\operatorname{sim} \left(\mathbf{u}_a^S \cdot \mathbf{u}_b^S\right) / \tau\right)},
	\end{aligned}
	\label{L-T}
\end{equation}
where $V^{T}$ is the set of nodes of the last snapshot graph $\mathcal{G}^{T}$.

\subsection{Output Layer}
The embedding $\mathbf{u}_a^T$ for node $a$ in the last snapshot $\mathcal{G}^T$ is utilized in the link prediction task, aiming to predict the existence of links between node $a$ and other nodes. Thus, this task is transformed into a similarity problem between node $a$ and its neighboring nodes in the last snapshot $\mathcal{G}^T$. We adopt binary cross-entropy minimization as our objective function, defined as follows:
\begin{equation}
	\mathcal{L}_{ {main}}=-\sum_{a\in V^{T}} \sum_{(a,i)\in \mathcal{O}^{+}}\log (\sigma({\mathbf{u}_{a}^{T}}^{\top}\cdot \mathbf{u}_{i}^{T} )) - \sum_{(a,j)\in \mathcal{O}^{-}}\log (\sigma({\mathbf{u}_{a}^{T}}^{\top}\cdot \mathbf{u}_{j}^{T} )), 
	\label{L-MIAN}
\end{equation}
where $\mathbf{u}_{a}^{T}$ for any node $a$ is defined as the mean pooling of $\mathbf{u}_{a}^{L}$ and $\mathbf{u}_{a}^{S}$, i.e., $\mathbf{u}_{a}^{T} = (\mathbf{u}_{a}^{L} + \mathbf{u}_{a}^{S}) / 2$. Within the graph $\mathcal{G}^{T}$, neighbors of node $a$ are defined as positive examples, while a random sample of non-neighbor nodes serves as negative examples, forming the triple $(a,i,j)$. The set $\mathcal{O}$ is defined as $\{(a,i,j)\}$. Correspondingly, the positive tuple is $\mathcal{O}^{+}=\{(a,i)\}$ and the negative tuple is $\mathcal{O}^{-}=\{(a,j)\}$.  

Then the total loss $\mathcal{L}_{total}$ is formulated by weighting three specific types of losses: the main cross-entropy loss $\mathcal{L}_{ {main}}$, node-level heterogeneity differentiation loss denoted by $\mathcal{L}_N = \mathcal{L}_{ {Node }}^{ {+ }}- \mathcal{L}_{\text {Node }}^{  {-}}$, edge-level heterogeneity differentiation loss expressed as $\mathcal{L}_E =  \mathcal{L}_{ {Edge }}^{  {+ }}-  \mathcal{L}_{  {Edge}}^{  {- }}$, and time-level heterogeneity differentiation loss represented by $\mathcal{L}_T = \mathcal{L}^{L+}_{ {Time }} + \mathcal{L}^{S+}_{ {Time }}$. 

\begin{equation}
	\mathcal{L}_{total} = \mathcal{L}_{ {main }}+\lambda_1 \mathcal{L}_N  +\lambda_2\mathcal{L}_E   +\lambda_3 \mathcal{L}_T,
	\label{L-TOTAL}
\end{equation}
where $\lambda_1$, $\lambda_2$, and $\lambda_3$ represent the learnable weighting factors employed to balance three losses.
The procedure of our \mymodel is outlined in Algorithm \ref{algorithm} in the Appendix. We define $d$, $N$, and $T$ as the node embedding dimension, the number of nodes, and snapshots, respectively. The time complexity of our \mymodel is $O(TNd^2)$.

\section{Experiments and Analysis}
\label{Sec:Experimental}

We conduct extensive experiments and compare our results with eight baselines across four datasets to investigate the following three research questions:
\begin{itemize}[leftmargin=*]
	\item \textbf{Q1:} How does \mymodel perform compared to state-of-the-art models?
	
	\item \textbf{Q2:} What role do key components in \mymodel play in enhancing its performance?
	
	
	\item \textbf{Q3:} How do hyperparameters affect the \mymodel's performance?
\end{itemize}

We first provide a concise overview of the experimental setup, followed by the responses to the aforementioned research questions.

\begin{table*}[!ht]
	\centering
	\caption{Overall Performance Comparisons. (The top two performances are highlighted, with the best boldfaced and the second-best underlined.)}
		\begin{tabular}{cccccllll}
			\toprule
			\multirow{2}{*}{Model}           & \multicolumn{2}{c}{Math-overflow}                    & \multicolumn{2}{c}{Taobao}                  & \multicolumn{2}{c}{OGBN-MAG}                         & \multicolumn{2}{c}{COVID-19}                        \\
			\cmidrule(r){2-3} \cmidrule(r){4-5} \cmidrule(r){6-7} \cmidrule(r){8-9}
			& AUC                  & AP                   & AUC                  & AP                   & \multicolumn{1}{c}{AUC} & \multicolumn{1}{c}{AP} & \multicolumn{1}{c}{AUC} & \multicolumn{1}{c}{AP} \\
			\midrule
			SEAL~\cite{zhang2018link}                                & 63.33                & 63.68                & 51.26                & 55.11                & 69.66                   & 74.74                  & 60.12                   & 63.62                  \\
			VGNAE~\cite{ahn2021variational}                             & 62.65                & 57.06                & 58.40                & 54.80                & 56.85                   & 53.72                & 72.89                 & 65.59                  \\
			Metapath2Vec~\cite{dong2017metapath2vec}                     & 64.62                & 72.30                 & 50.20                 & 66.48                & 62.32                   & 74.11                  & 54.66                   & 67.32                  \\
			GATNE~\cite{cen2019representation}                            & 71.53                & 73.87                & 71.32                & 82.71                & 68.01                   & 77.63                  & 71.15                   & 66.30                   \\
			TGAT~\cite{XuRKKA20}                             & 66.46                & 60.25                & 77.49                & 71.38                & 61.72                   & 63.06                  & 61.44                   & 64.62                  \\
			TDGNN~\cite{qu2020continuous}                             & 83.09                & 74.69                & 71.33                & 73.84               & 78.94                   & 70.66                  & 67.28                   & 65.77                   \\
			THAN~\cite{li2023memory}                               & {\underline{89.23} }          & {\underline{89.05} }               & 65.22                & 65.51                & 74.63                   & 71.88                  &   {\underline{ 74.07}}             & {\underline{ 70.49}}            \\
			THGAT~\cite{zhang2023dynamic}                            & 77.27                &  80.88          & {\underline{80.04}}          & {\underline{ 84.30}}           & {\underline{ 85.97}}             & {\underline{ 86.42}}            & 64.88                   & 67.62                  \\
			\textbf{\mymodel} & \textbf{92.34}       & \textbf{91.06}       & \textbf{80.44}       & \textbf{92.82}       & \textbf{88.55}          & \textbf{94.27}         & \textbf{77.96}             & \textbf{83.98}         \\
			\bottomrule
		\end{tabular}
	\label{tab:baselines}
\end{table*}

\subsection{Experimental Setup}
\subsubsection{Datasets and Metrics}
We perform experiments on four application scenarios, i.e., Math-overflow, Taobao, OGBN-MAG, and COVID-19, to verify the universality and effectiveness of our proposed \mymodel.
We employ widely-adopted Area Under the Curve (AUC) and AP as evaluation metrics~\cite{li2023memory,XuRKKA20}.
The specifics and processing details of four datasets are presented in $\S$ \ref{sec:dataset_metric_appendix} of the Appendix.


\subsubsection{Baselines}
We evaluate our \mymodel against eight baselines categorized into four groups:
(1) Static Homogeneous(SEAL~\cite{zhang2018link}, VGNAE~\cite{ahn2021variational}),
(2) Static Heterogeneous (Metapath2Vec~\cite{dong2017metapath2vec}, GATNE~\cite{cen2019representation}),
(3) Dynamic Homogeneous (TGAT~\cite{XuRKKA20}, TDGNN~\cite{qu2020continuous}), and
(4) Dynamic Heterogeneous (THAN~\cite{li2023memory}, THGAT~\cite{zhang2023dynamic}).
Further detailed elaboration on baseline models and our model is available in $\S$ \ref{sec:baseline_appendix} of the Appendix.

\subsubsection{Implementation Details}

We have made the source code and datasets publicly available at \href{https://anonymous.4open.science/r/CLP-7E66}{https://anonymous.4open.science/r/CLP-7E66}. During the training process, our \mymodel is executed with a batch size of 1024.
We employ an early-stopping strategy halting training when the Average Precision (AP) metric ceases to increase for $5$ consecutive epochs. 
The learning rate $lr$ is set at 1e-4.
We employ Adam as the optimizer of our CLP model.
The hyper-parameters are carefully optimized following a grid search.
More details can be found in $\S$ \ref{sec:implemention_appendix} of the Appendix.

\subsection{Performance and Analysis}

In order to address the aforementioned three questions \textbf{Q1}-\textbf{Q3}, we execute the following experiments and analyze the respective results.

\subsubsection{Overall Performance Comparisons (for \textbf{Q1)}}

To answer \textbf{Q1}, we perform a comparative evaluation of our \mymodel by comparing it against eight baselines on Math-overflow, Taobao, OGBN-MAG, and COVID-19 datasets. The results are presented in Table \ref{tab:baselines}. Therefore, we obtain the following observation analyses.

(1) The static homogeneous networks, SEAL and VGNAE, exhibit inadequate performance across all four datasets, primarily owing to their inability to leverage temporal dynamics and heterogeneous information. SEAL is specifically designed for homogeneous networks and struggles with handling multiple types of nodes and edges. It relies on subgraph extraction and graph neural networks, which makes real-time updates challenging. As for VGNE, lack adaptability to network changes, which hampers maintaining efficiency in dynamic environments.

(2) The static heterogeneous networks, Metapath2Vec and GATNE, achieve superior performance over SEAL and VGNAE by exploiting heterogeneous information. Specifically, Metapath2Vec meticulously designs meta-paths for heterogeneous networks, while GATNE effectively integrates both the topological heterogeneity and node attribute information. However, both models overlook the network dynamics, which can lead to inaccurate node representations, as connections may change from negative to positive between training and testing phases.

(3) The dynamic homogeneous networks, TGAT and TDGNN, excel by encoding temporal-topological features, significantly outperforming the aforementioned SEAL, VGNAE, Metapath2Vec, and GATNE. However, TGAT focuses only on temporal-topological nodes and time-feature edges, neglecting the diversity of nodes and edges. On the other hand, TDGNN emphasizes only the temporal aspects of edges in node representations, disregarding edge multiplicity. Therefore, the effectiveness of TGAT and TDGNN is limited when compared to THAN and THGAT.

(4) The dynamic heterogeneous networks, THAN and THGAT, exhibit robust performance through advanced modeling techniques. For THAN, it captures sufficient heterogeneous information and builds continuous dynamic relationships through the analysis of temporal causality. For THGAT, it designs a node signature method tailored to heterogeneous data and incorporates a temporal heterogeneous graph attention layer, effectively integrating both heterogeneous and temporal information. These two models effectively model both temporal dynamics and heterogeneous semantics in graph data, as evidenced by superior performance compared to other baseline models.

(5) Our model, \textbf{\mymodel}, markedly surpasses all baseline models in terms of AUC and AP across four datasets. Compared to the second best model, THGAT, which relies on a coarse-grained view to capture heterogeneous and temporal information, our \mymodel employs a fine-grained perspective to effectively encode the spatial heterogeneity and temporal heterogeneity. Initially, we devise a heterogeneous temporal graph to delineate the underlying structural distribution patterns and sequential evolution paradigms. Subsequently, we propose a contrastive hierarchical discrepancy elimination module to incorporate intrinsic inter-relations at node-, edge-, and time-level, respectively. As a result, our model achieves significant performance enhancements over THGAT, with average increases of 10.10\% and 13.44\% in terms of AUC and AP, respectively.

(6) The performance of all models on the COVID-19 dataset exhibits a general decline. This downturn can be attributed to the scarcity of observational nodes, which contributes to data instability and suboptimal link prediction results. Moving forward, we will strive to enhance our model's performance and intensify our modeling efforts to better understand complex patterns and dynamics associated with disease transmission pathways.

\begin{figure}[!t]
	\scriptsize
	\centering
	\subfigure[{AP}]
	{
		\includegraphics[width=.4\linewidth]{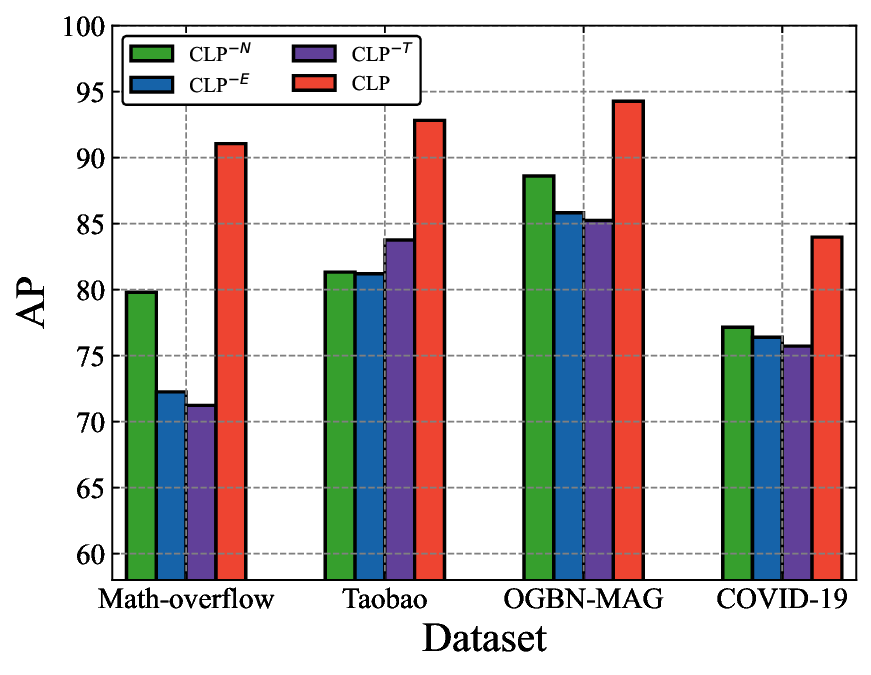}
		\label{subfig:Weibo-MSLE}
	}
	\subfigure[{AUC}]
	{
		\includegraphics[width=.4\linewidth]{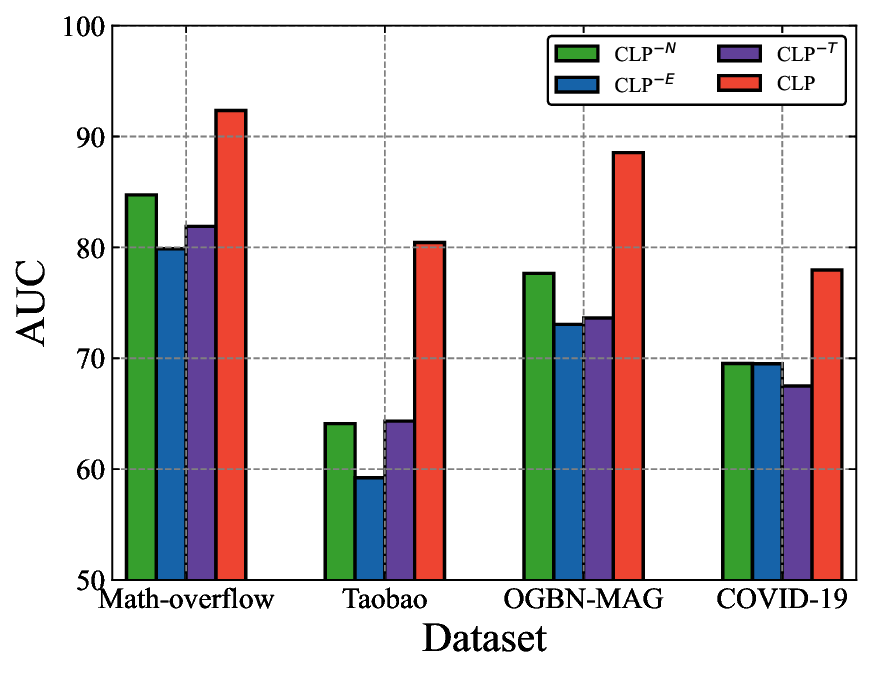}
		\label{subfig:Weibo-mSLE}
	}
	\caption{Ablation Experimental Results.}
	\label{Fig:ablation}
\end{figure}

\subsubsection{ Ablation Experiments (for \textbf{Q2})}
For \textbf{Q2}, we perform ablation studies on four datasets to evaluate the effectiveness of the core components in \mymodel.
Our analyses primarily focus on the ablation of node, edge, and time-level loss functions in the hierarchical heterogeneity differentiation modeling, designated as $\mathbf{\clp}^{-N}$, $\mathbf{\clp}^{-E}$, and $\mathbf{\clp}^{-T}$, respectively.
Figure~\ref{Fig:ablation} displays the comparison results.
These findings indicate that our proposed \mymodel performs significantly better than its variant models.

(1) \textbf{$\mathbf{\clp}^{-E}$ exhibits the most substantial performance degradation.}
Specifically, removing the edge-level heterogeneity differentiation loss leads to a marked decrease in predictive accuracy, with average reductions of 19.08\% in AUC and 12.67\% in AP. Notably, Taobao experiences a 35.65\% decrease in terms of AUC, underscoring the critical role of the edge-level discrepancy elimination module in our model.

(2) $\mathbf{\clp}^{-T}$ results in the second most significant performance drop. Specifically, when we eliminate the time-level heterogeneity differentiation loss, there is a considerable decrease in prediction performance, with average decreases of 16.23\% and 12.67\% in AUC and AP, respectively. This observation validates the vital importance of the time-level discrepancy elimination module in augmenting the overall node representation.

(3) $\mathbf{\clp}^{-N}$ also indicates a decline in link performance. Specifically, when we remove the node-level heterogeneity differentiation loss, there is a notable decrease in AUC and AP, with an average reduction of 13.05\% and 8.65\%, respectively. This result confirms the essential contribution of the node-level nonuniformity elimination module. 

In conclusion, the node-level, edge-level, and time-level heterogeneity differentiation modeling modules constitute essential components in \mymodel that fundamentally enhance its capability to predict temporal heterogeneous links accurately.

\subsubsection{Effect of Parameters (for \textbf{Q3})}
In response to \textbf{Q3}, we conduct a sensitivity analysis of the key parameters of our model on Taobao and OGBN-MAG. These parameters include the dimension of vector representation  ($d$), the number of multi-heads ($h$), loss weights ($\lambda_1$, $\lambda_2$, and $\lambda_3$) and the temperature coefficient ($\tau$). 

\begin{figure}
	\scriptsize
	\centering
	\subfigure[\scriptsize{Node Vector Dimension w.r.t. AP}]
	{
		\includegraphics[width=.3\linewidth]{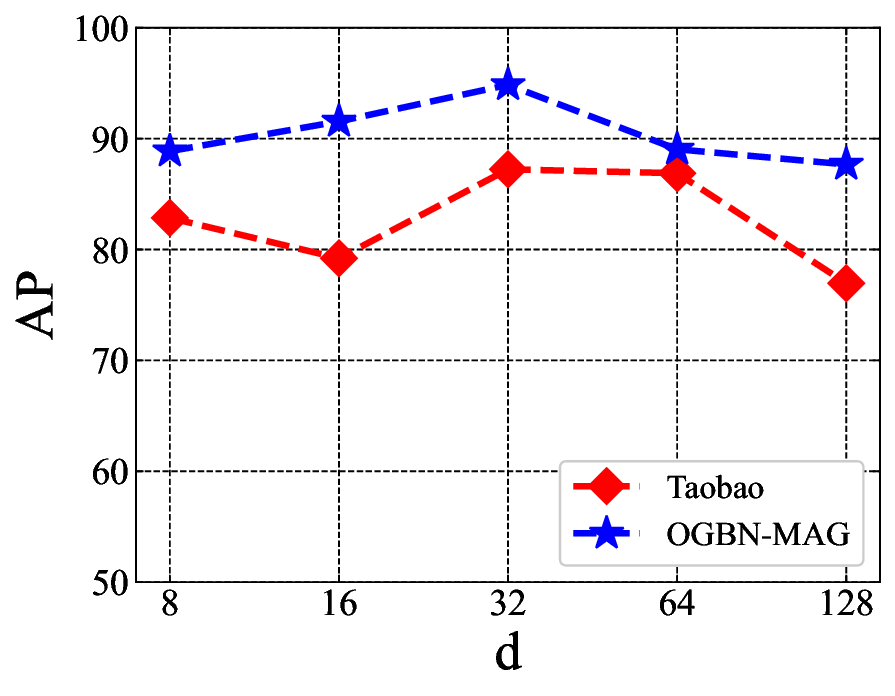}
		\label{subfig:Dim FOR AP}
	}
	\qquad
	\subfigure[\scriptsize{Node Vector Dimension w.r.t. AUC}]
	{
		\includegraphics[width=.3\linewidth]{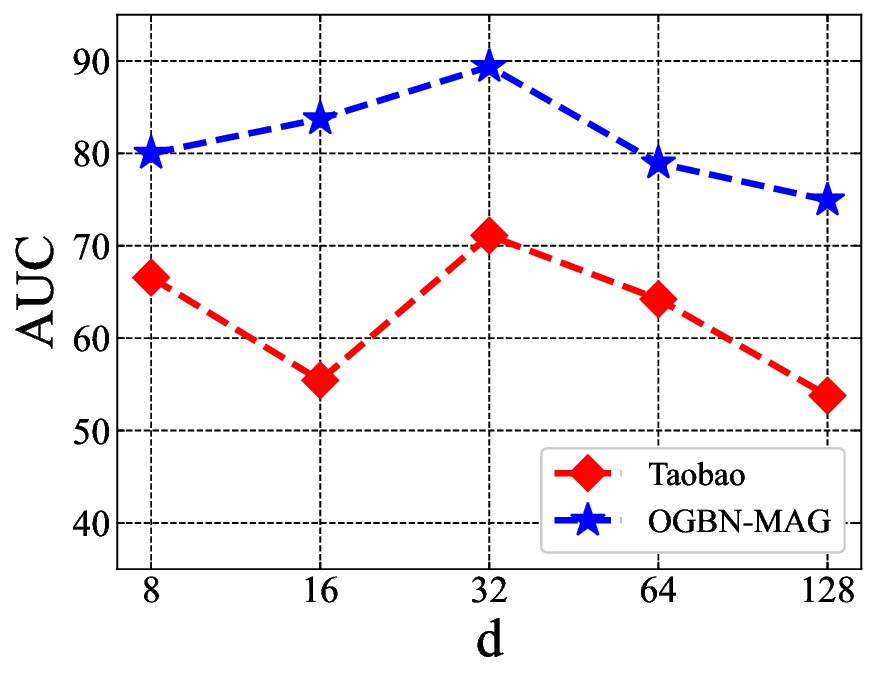}
		\label{subfig:Dim FOR AUC}
	}
	\caption{Impact of $d$.}
	\label{Fig:d}
\end{figure}

(1) \textbf{Dimension of representations} ($d$): 
We search the optimal value of $d$ in the range of $\left\{8, 16, 32, 64, 128\right\}$, as shown in Figure~\ref{Fig:d}. 
As $d$ increases, \mymodel demonstrates enhanced performance, primarily due to the increased capacity and complexity of its learnable parameters. This expansion enables our model to capture more nuanced data features, significantly improving accuracy and analytical depth. However, excessive dimensions can lead to a decline in model performance, attributable to overfitting and noise in vector representations. Such disadvantages arise because larger vector spaces not only capture essential features but also irrelevant data variations, which can obfuscate the learning process and degrade generalization on new datasets. Empirical evidence from our experiments identifies $d=32$ as the optimal setting for \mymodel, striking a balance between accuracy and robustness. This setting prevents overfitting and effectively captures essential patterns necessary for prediction and analysis.

\begin{figure}
	\scriptsize
	\centering
	\subfigure[\scriptsize{Number of heads w.r.t. AP}]
	{
		\includegraphics[width=.3\linewidth]{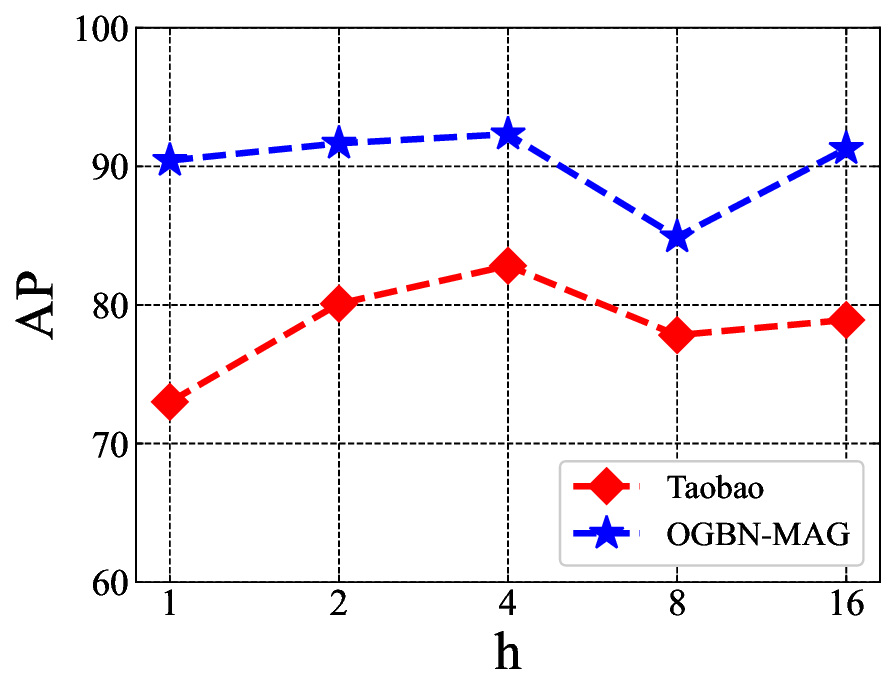}
		\label{subfig:HEAD FOR AP}
	}
	\qquad
	\subfigure[\scriptsize{Number of heads w.r.t. AUC}]
	{
		\includegraphics[width=.3\linewidth]{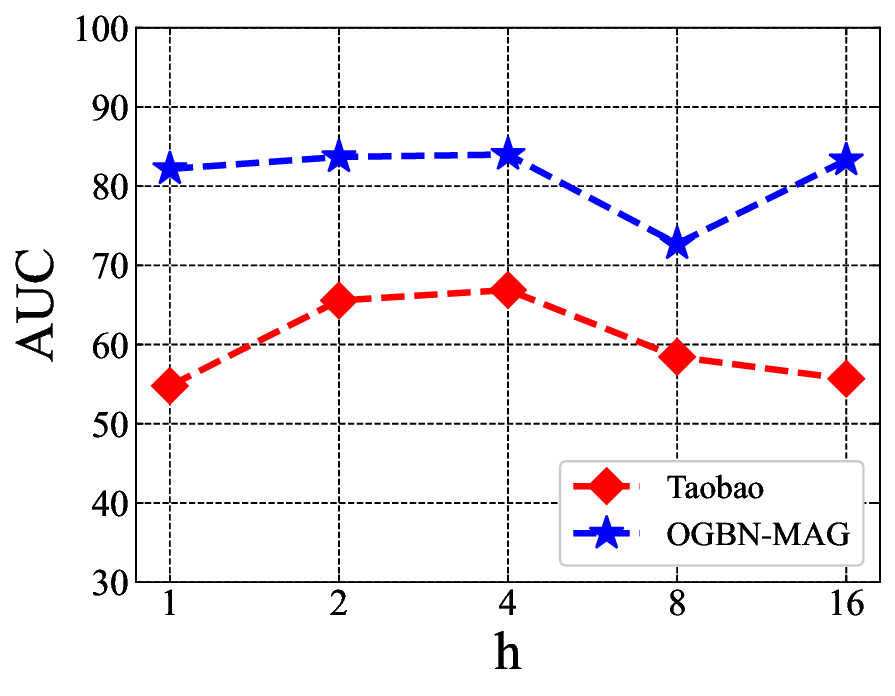}
		\label{subfig:HEAD FOR AUC}
	}
	\caption{Impact of $h$.}
	\label{Fig:h}
\end{figure}

(2) \textbf{Number of heads} ($h$): 
The $h$-head attention mechanism segregates sub-semantic spaces at the node-level and edge-level feature learning module, enabling our model to direct its attention towards different heterogeneous information dimensions. Adjusting $h$ allows our \mymodel to tailor its focus to specific aspects of the input data, thereby enhancing its feature extraction and representation capabilities. As depicted in Figure \ref{Fig:h}, the optimal expressive ability and attention allocation capability of our model are achieved when $h=4$. This configuration promotes a balance between computational efficiency and model performance, facilitating precise management of the complexities involved in diverse data interactions. 

\begin{figure}
	\scriptsize
	\centering
	\subfigure[$\lambda_1$ w.r.t. AP]
	{
		\includegraphics[width=.31\linewidth]{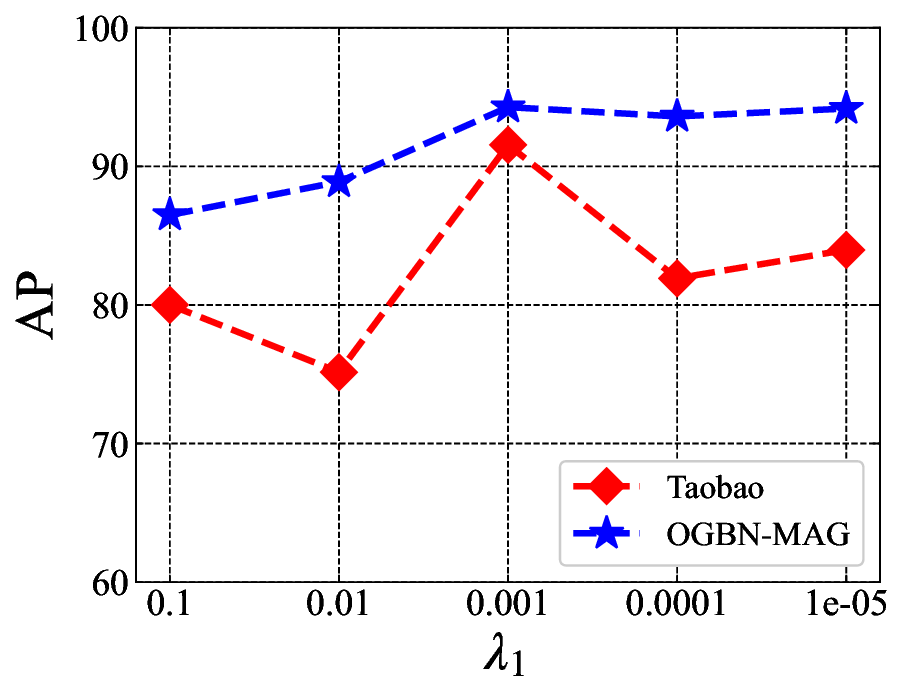}
		\label{subfig: LAM1 FOR AP}
	}
	\subfigure[$\lambda_1$ w.r.t. AUC]
	{
		\includegraphics[width=.31\linewidth]{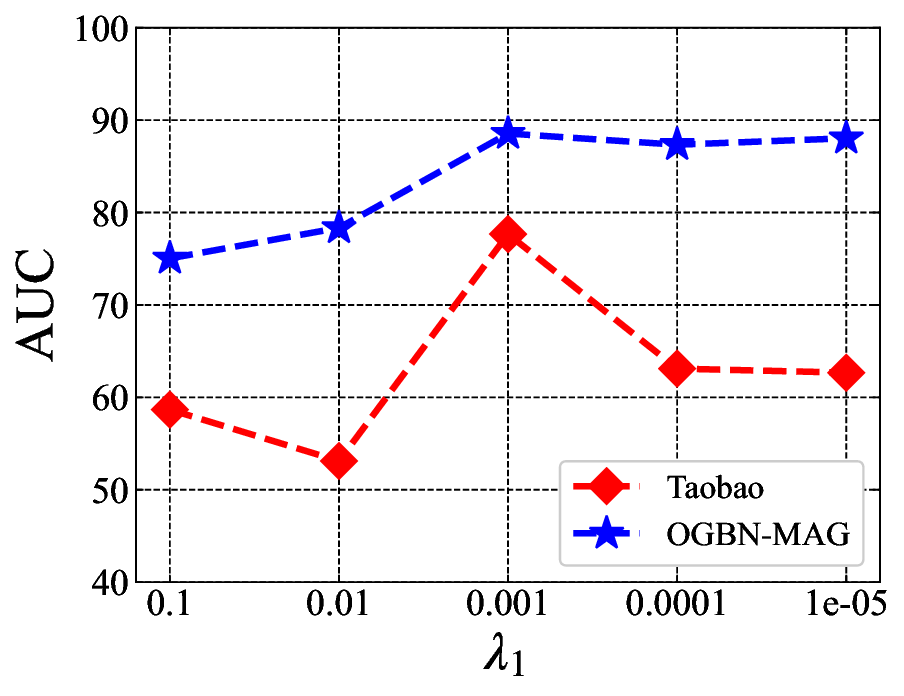}
		\label{subfig:LANM1 FOR AP}
	}
	\subfigure[$\lambda_2$ w.r.t. AP]
	{
		\includegraphics[width=.31\linewidth]{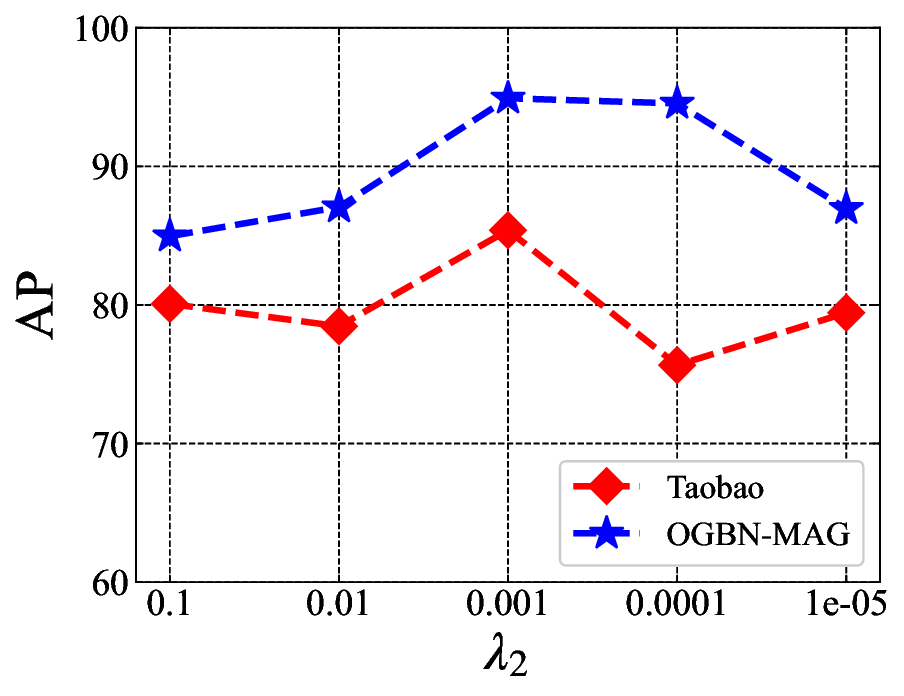}
		\label{subfig:LANM2 FOR AP}
	}
	\subfigure[$\lambda_2$ w.r.t. AUC]
	{
		\includegraphics[width=.31\linewidth]{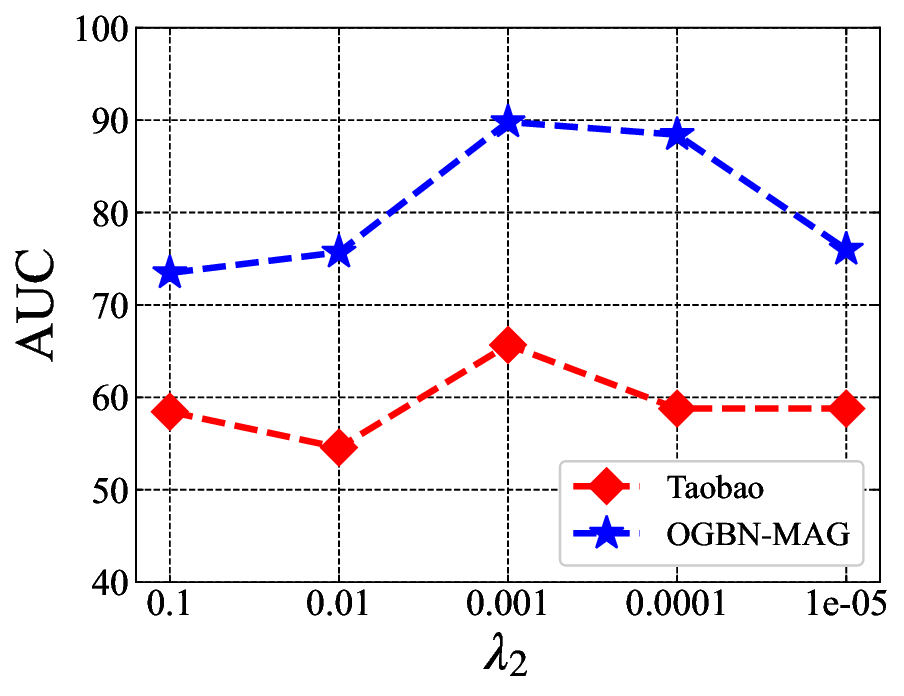}
		\label{subfig:LANM2 FOR AUC}
	}
	\subfigure[$\lambda_3$ w.r.t. AP]
	{
		\includegraphics[width=.31\linewidth]{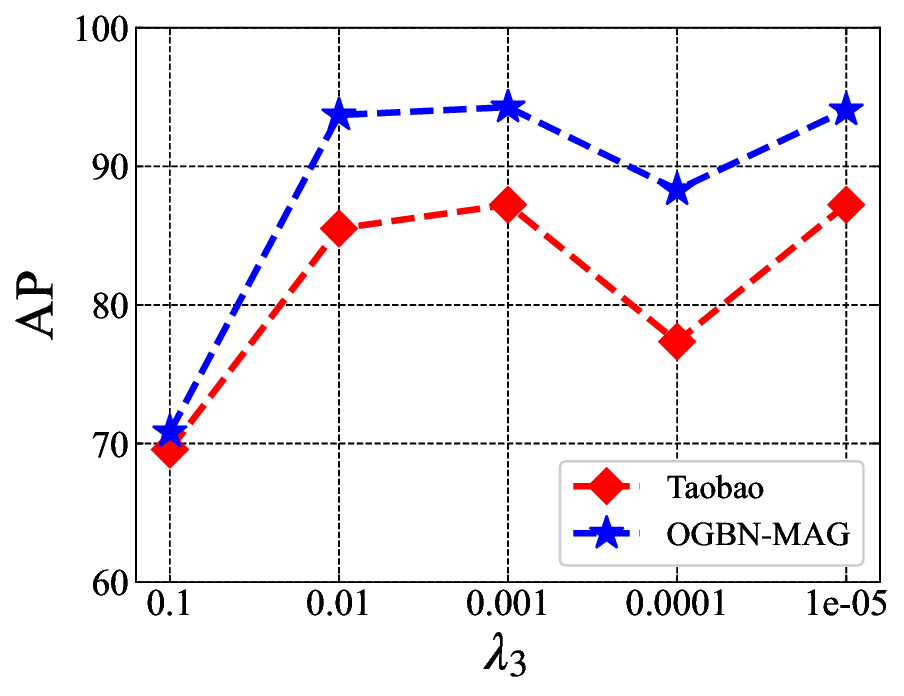}
		\label{subfig:LANM3 FOR AP}
	}
	\subfigure[$\lambda_3$ w.r.t. AUC]
	{
		\includegraphics[width=.31\linewidth]{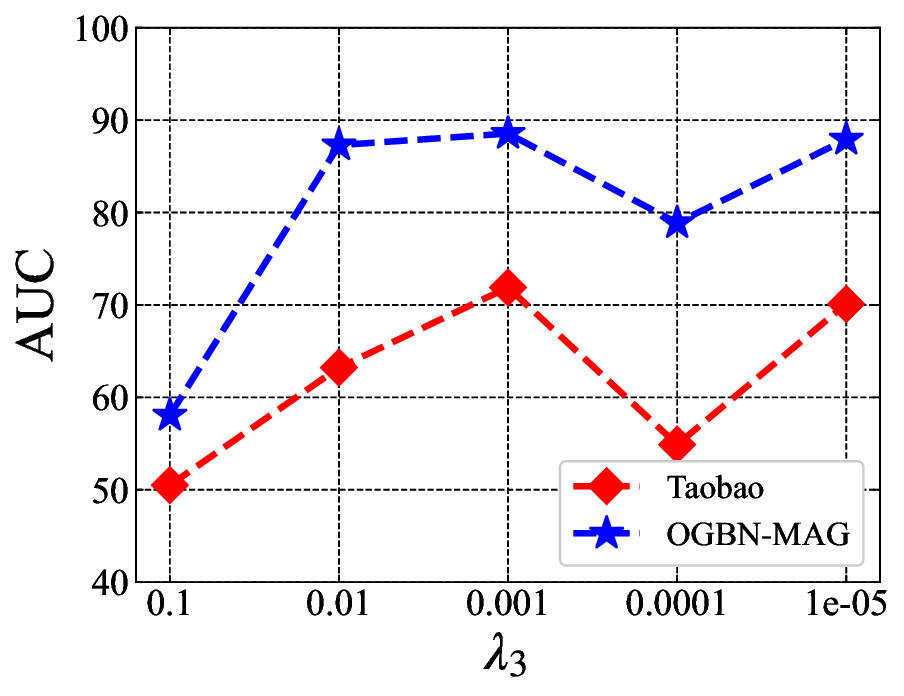}
		\label{subfig:LANM3 FOR AUC}
	}
	\caption{Impact of $\lambda_1$, $\lambda_2$, and $\lambda_3$.}
	\label{Fig:λ}
\end{figure}

(3) \textbf{Loss weights} ($\lambda_1$, $\lambda_2$, and $\lambda_3$): 
Adjusting the weights for node-, edge-, and time-level contrastive heterogeneity differentiation loss is critical for balanced optimization during model training. Precise tuning of these parameters ensures that each component contributes proportionally to its role in processing heterogeneous data, enhancing overall model performance. 
The optimal prediction performance is achieved when $\lambda_1$, $\lambda_2$, and $\lambda_3$ are all set to $1e-3$, as depicted in Figure~\ref{Fig:λ}. This setting notably minimizes errors in the final output by effectively balancing the emphasis on different types of losses, customizing our \mymodel to link prediction task, pivotal in achieving high performance in dynamic heterogeneous scenarios.

\begin{figure}[!t]
	\centering
	\scriptsize
	\subfigure[$\tau$ w.r.t. AP]
	{
		\includegraphics[width=.3\linewidth]{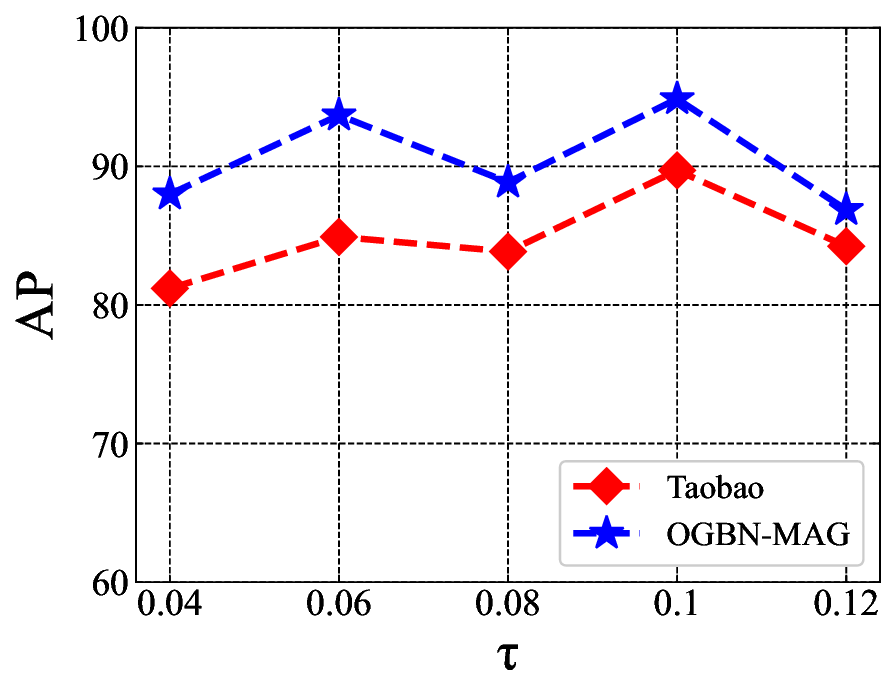}
		\label{subfig:tao for AP}
	}
	\qquad
	\subfigure[$\tau$ w.r.t. AUC]
	{
		\includegraphics[width=.3\linewidth]{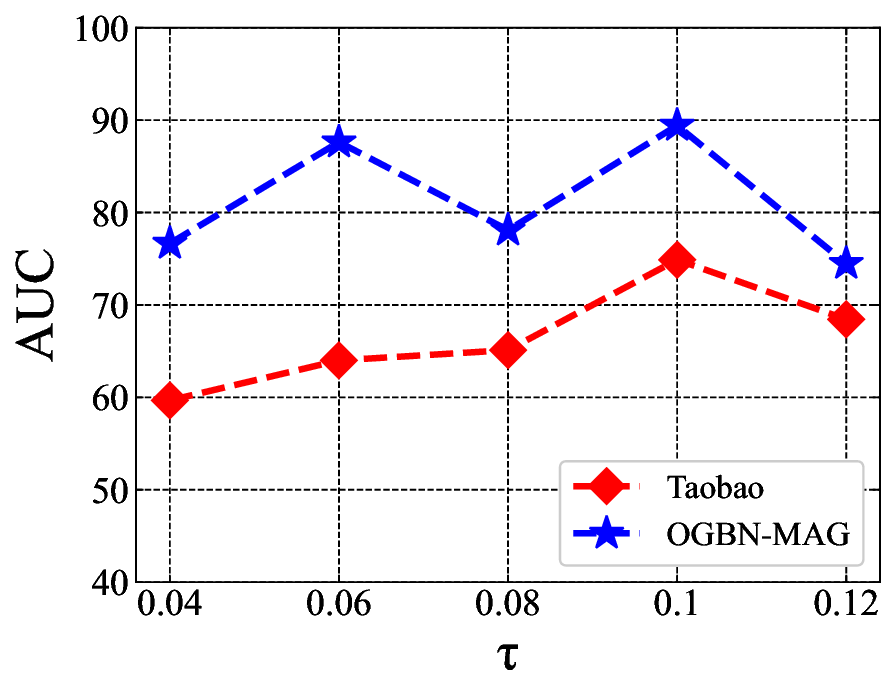}
		\label{subfig:tao for AUC}
	}
	\caption{Impact of $\tau$.}
	\label{Fig:tau}
\end{figure}

(4) \textbf{Temperature coefficient ($\tau$)}: $\tau$ modulates the sensitivity to sample similarities during training, impacting overall model efficacy. 
Specifically, optimal $\tau$ value enables the contrastive loss function to accurately differentiate between similar and dissimilar samples. As shown in Figure~\ref{Fig:tau}, the value of $0.1$ yields the most effective outcomes, significantly optimizing contrastive learning performance within our \mymodel. This setting encourages the model's focus on meaningful variances among samples, thereby improving both accuracy and robustness in real-world scenarios. 

\section{Related Work}\label{Sec:Related Works}


\subsection{Heterogeneous Network Embedding Approaches}
\label{Heterogeneous graph}
Heterogeneous network embedding approaches focus on capturing structural and semantic features in complex networks with diverse node and edge types to improve link prediction accuracy~\cite{wang2022survey}.
Some studies \cite{dong2017metapath2vec, fu2017hin2vec, wang2019heterogeneous, qiu2018network, zhao2023link, zhang2020mg2vec} employ meta-paths and meta-graphs for efficient sampling of heterogeneous information.
However, these methods may overlook node and edge attributes.
To overcome such limits, some researches, including \cite{hu2020heterogeneous, li2020type, nguyen2023link, zhang2023page}, pay their attention to aggregating diverse neighbor information, while others~\cite{xu2021topic,peng2023th} focus on the topic-aware node attributes. 
However, these methods rely on the accurate integration of diverse data types, and lack the in-depth inspection of data distribution.

\subsection{Temporal Network Embedding Approaches}
\label{Temporal graph}

Temporal graph learning approaches can be divided into sequential and graph models to handle the dynamic nature of real-world networks. Sequential models~\cite{hao2020dynamic,chen2019lstm,jiang2023dnformer,zhang2023attentional} divide the THNs into a series of temporal snapshots and employ RNNs~\cite{hao2020dynamic,chen2019lstm} and attention mechanisms~\cite{jiang2023dnformer,zhang2023attentional} to learn snapshot dependencies. Meanwhile, graph models~\cite{lei2019gcn,yang2019advanced,pareja2020evolvegcn} are better suited for handling variable-length sequence data and adapting to the fluctuating nature of time-series data compared to sequence models. Specifically, models including~\cite{lei2019gcn,yang2019advanced,pareja2020evolvegcn,XuRKKA20} leverage GNNs to capture evolving relationships and topological properties within dynamic networks. Other researches~\cite{pham2021comgcn,yang2022few,fan2022heterogeneous,luo2023graph,huang2023temporal,zhang2023dynamic1} further enhance dynamic network embedding through various aggregation and modeling strategies.
However, despite the advantages of these approaches in capturing temporal dynamics, challenges remain in terms of computational demands.

\subsection{Contrastive Learning Approaches}
\label{Contrastive learning}
Contrastive learning~\cite{chen2020simple,LiuJPZZXY23,WuLTGL23} is a promising approach for link prediction, utilizing unsupervised representation learning with positive and negative sample pairs. Notable work like GraphCL~\cite{you2020graph} employs random graph augmentations to generate pairs of graph instances for improved representation learning. HeCo~\cite{WangLHS21} introduces a co-contrastive mechanism preserving both meta-path and network schema information effectively. LGCL~\cite{zhang2023line} transforms link prediction into node classification by using line graphs and applying contrastive learning to maximize mutual information. SMiLE~\cite{PengLXXWP22} focuses on incomplete graphs by employing a dual-view encoder to capture structural and contextual information for entity representation. Studies in~\cite{XuDXJ23} differentiate between groups based on structural similarity rather than individual nodes, optimizing their embeddings accordingly. 
Nevertheless, these methods enhance link prediction models' discriminative power but often overlook fine-grained nonuniformity.

\section{Conclusion}
\label{sec:Conclusion}
In this paper, we propose a novel contrastive learning-based link prediction method, \textbf{\mymodel}, which develops a multi-view hierarchical self-supervised architecture to effectively characterize spatial and temporal heterogeneity. First, for spatial heterogeneity, we introduce a spatial feature modeling component to identify topological distribution patterns at node- and edge- level. Next, for temporal heterogeneity, we devise a dual-channel architecture that represents the evolutionary progressions within dynamic graph topologies over both the long and short terms. Lastly, we design a multi-view hierarchical representation architecture to integrate the heterogeneity paradigms, effectively encoding spatial and temporal distribution complexity. The efficacy of our model for temporal heterogeneous link prediction is validated through experiments on four public datasets, demonstrating its significantly superiority over state-of-the-art methods in link prediction tasks.
Our future research aims to extend the application of THNs to diverse domains for advanced network analysis and prediction. Particularly, we plan to incorporate multimedia data, such as images and videos, into THNs to enrich our exploration and comprehension of multimedia content within these networks. By utilizing the dynamic features of THNs, we can delve into complex pattern recognition and anomaly detection in multimedia streams.

\begin{acks}
	This work was supported in part by the National Key Research and Development Program of China (2020YFB1406902), the Key-Area Research and Development Program of Guangdong Province (2020B0101360001), the Shenzhen Science and Technology Research and Development Foundation (JCYJ20190806143418198), the Major Key Project of PCL (Grant No. PCL2021A02), and the Fundamental Research Funds for the Central Universities (HIT.OCEF.2021007). Professor Hui He (hehui@hit.edu.cn) and Professor Weizhe Zhang (wzzhang@hit.edu.cn) are the corresponding authors.
\end{acks}


\bibliographystyle{ACM-Reference-Format}
\bibliography{CLP}

\clearpage

\appendix
\section*{Appendix}

\section{Algorithm}

The detailed algorithm of our \mymodel is presented in Algorithm \ref{algorithm}.

\renewcommand{\algorithmicrequire}{ \textbf{Input:}}     
\renewcommand{\algorithmicensure}{ \textbf{Output:}}    

\begin{algorithm}
	\caption{Link Prediction with \mymodel.}
	\label{algorithm}
	\begin{algorithmic}[1]
		\REQUIRE The sequence of graph snapshots: $\mathcal{G}=\{\mathcal{G}^1, \mathcal{G}^2, \cdots, \mathcal{G}^T\}$.
		\ENSURE Final node representation embedding: $\mathbf{u}_a$.
		\FOR{Each train iteration}
		\STATE  Parameters initialization.
		
		\FOR{$\mathcal{G}^t$ in $\mathcal{G}$}
		\STATE Construct subgraph $\mathcal{G}^{rt}$ based on the edge type $r \in R$;
		\FOR{Every subgraph $\mathcal{G}^{rt}$ associated with edge type $r$}
		\STATE Calculate the node-level weight $\alpha_{a,b}^{rt}$ for any node pair $(a,b)$ via Eq.~\eqref{N-S} and Eq.~\eqref{N-W};
		\STATE Obtain the node-level heterogeneous embedding $\mathbf{u}_a^{rt}$ for any node $a$ via Eq.~\eqref{N-H1};
		\STATE Obtain the node-level unified embedding $\mathbf{h}_a^{rt}$ for any node $a$ via Eq.~\eqref{N-H2};
		\STATE Calculate the node-level loss $\mathcal{L}_{Node}^+$ and $\mathcal{L}_{Node}^-$ via Eq.~\eqref{loss-N};
		\ENDFOR	 
		\STATE Calculate the edge-level weight $\delta_{a}^{r t}$ for any node $a$ via Eq.~\eqref{edge-s};
		\STATE Obtain the edge-level heterogeneous embedding $\mathbf{u}_a^{t}$ for any node $a$ via Eq.~\eqref{E-H1};
		\STATE Obtain the edge-level unified embedding $\mathbf{h}_a^{t}$ for any node $a$ via Eq.~\eqref{E-H2};
		\STATE Calculate the edge-level loss $\mathcal{L}_{Edge}^+$ and $\mathcal{L}_{Edge}^-$ via Eq.~\eqref{loss-E};   	
		\ENDFOR	
		\STATE Obtain temporal long- and short-term representations $\mathbf{u}_a^L$ and $\mathbf{u}_a^S$ via Eq.~\eqref{T-L} and Eq.~\eqref{T-G};
		\STATE Obtain node representation embedding: $\mathbf{u}_a$ through the mean pooling of $\mathbf{u}_a^L$ and $\mathbf{u}_a^S$;
		\STATE Calculate the time-level loss $\mathcal{L}_{Time}^{L+}$ and $\mathcal{L}_{Time}^{S+}$ via Eq.~\eqref{L-T};    	
		\STATE Calculate the total loss $\mathcal{L}_{total}$ via Eq.~\eqref{L-MIAN} and Eq.~\eqref{L-TOTAL}.    		
		\ENDFOR
		\RETURN{$\mathbf{u}_a$ }		
	\end{algorithmic}
\end{algorithm}

\begin{table*}[t]
	\caption{Comparisons of Baseline Models and Our Proposed Approaches.}
	\scriptsize
		\begin{tabular}{cccc}
			\toprule
			\multicolumn{3}{c}{Model}                                                                               & \multicolumn{1}{c}{Key Components}                           \\ \midrule
			& \multirow{2}{*}{Static Homogeneous}     & SEAL~\cite{zhang2018link}                             & Node Labeling, Subgraph Extraction, GNN Graph Learning       \\
			\multirow{7}{*}{Baselines} &                                         & VGNAE~\cite{ahn2021variational}                           & Graph Autoencoder, Variational Inference, GCN Graph Learning \\
			& \multirow{2}{*}{Static   Heterogeneous} & Metapath2Vec~\cite{dong2017metapath2vec}                     & Metapath-guided Random Walks, Skip-gram Modeling             \\
			&                                         & GATNE~\cite{cen2019representation}                            & Attributed Multiplex Networks, GCN Graph Learning            \\
			& \multirow{2}{*}{Dynamic   Homogeneous}  & TGAT~\cite{XuRKKA20}                              & Graph Attention Network, Self-attention Mechanism             \\
			&                                         & TDGNN~\cite{qu2020continuous}                             & GCN Graph Learning, Temporal Aggregator, Edge Aggregator     \\
			& \multirow{3}{*}{Dynamic Heterogeneous}  & THAN~\cite{li2023memory}                              & Dynamic Transfer Matrix, Self-attention Mechanism             \\
			&                                         & THGAT~\cite{zhang2023dynamic}                            & Neighborhood Type Modeling and Aggregation, Temporal Dynamics Integration                 \\
			\multirow{4}{*}{\textbf{Ours}} &                                         & \textbf{\mymodel} & Node-level Feature Modeling, Edge-level Feature Modeling, Temporal Information Modeling                         \\
			& \multicolumn{1}{l}{}                    & \textbf{\mymodel}$^{-N}$ & Edge-level Feature Modeling, Temporal Information Modeling                 \\
			& \multicolumn{1}{l}{}                    & \textbf{\mymodel}$^{-E}$ &Node-level Feature Modeling, Temporal Information Modeling                        \\
			& \multicolumn{1}{l}{}                    & \textbf{\mymodel}$^{-T}$ & Node-level Feature Modeling, Edge-level Feature Modeling         
			\\
			\bottomrule
		\end{tabular}
	\label{tab:baselineModel}
\end{table*}

\section{More Detailed Settings}

\subsection{Implemention Details}
\label{sec:implemention_appendix}

In this section, We provide more detailed fine-tuning of our hyper-parameters and baseline configurations.
Specifically, the dimension $d$ of node embeddings is explored within the range of $\left\{8, 16, 32, 64, 128\right\}$.
Moreover, we search for the optimal values of the balance coefficients $\lambda_1$, $\lambda_2$, and $\lambda_3$ in the range of $1e\left\{-1, -2, -3, -4, -5\right\}$; the number of attention heads $h$ in the range of $\left\{1, 2, 4, 8, 16\right\}$; and the temparature coefficient $\tau$ in the closed interval of $[0.04, 0.12]$ with a step size of $0.02$.
By default, after fine-tuning, we adopt the following hyperparameter settings, wherein the optimal values of $d, \lambda_{\cdot}, h, \text{ and } \tau$ are set to $32, 1e-3, 4, \text{ and } 0.1$, respectively.

For our baseline configuration, Metapath2Vec is set with a sequence walk length of 5, generating 10 walk sequences per node. The context size is 4, and the dimension of the node embedding vector is 32. The relevant parameters for SEAL, VGNAE, GATNE, TGAT, TDGNN, THAN, and THGAT are consistently maintained in accordance with their respective configurations.  In the case of homogeneous models such as SEAL, VGNAE, TGAT, and TDGNN, node type and edge type information is directly eliminated from the graph data during the experiments. For static models like SEAL, VGNAE, Metapath2Vec, and GATNE, we adopt the static graph representation learning approach, integrating edge data into a unified graph for comprehensive training. We implement and fine-tune baseline models using their official codes and adhere to the optimized setting values for all other hyperparameters of the baselines as reported in their respective papers.
All experiments are implemented on NVIDIA RTX A2000 (12G). 

\subsection{Datasets and Metrics}
\label{sec:dataset_metric_appendix}

This section will present the details and processing of the datasets, along with the evaluation metrics. The dataset statistics are presented in Table \ref{tab:datasets}.

\subsubsection{Dataset Description}

\begin{itemize}[leftmargin=*]
	\item \textbf{Math-overflow} \footnote{http://snap.stanford.edu/data/sx-mathoverflow.html.}: Math-overflow serves as an interactive website for mathematics enthusiasts. It adopts a forum-style model for communication and interaction, fostering an online community of expert mathematicians. This dataset comprises 2350 days of user interactions, which can be categorized into three types: question and answer exchanges, question and comment discussions, and answer and comment engagements. For analytical purposes in our experiments, we divide this dataset into 11 snapshots using a time window of 124 days.

	\item \textbf{Taobao}~\cite{DuWYZT19}: Taobao is characterized by three types of node: user (U), product (I), and topic (T), along with three types of links: U-I, U-T, and I-T, which signify users' interactions on cloud-themed products in the Taobao App from April 1 to May 31, 2008. For our experimentation, this dataset is divided into five snapshots using a time window of 12 days.
	
	\item \textbf{OGBN-MAG}~\cite{hu2020open}: OGBN-MAG is a subset of Microsoft Academic Graph (MAG), which contains four node types (papers, authors, institutions, and research areas) and four types of relationships (authors affiliated with institutions, authors writing papers, papers citing other papers, and papers associated with research areas as their topics). In our experiments, a THN is extracted from OGBN-MAG for the period from January 1st to 10th, 2010, divided into time slots with 1000 edges of each type to create the dataset for the experiment.
	
	\item \textbf{COVID-19}~\footnote{https://coronavirus.1point3acres.com/en}: COVID-19, sourced from 1point3acres, provides state- and county-level daily case reports, including confirmed cases, new cases, deaths, and recovered cases. We select the daily new COVID-19 cases as our THN data. This dataset consists of two node types: (state and county) and three relationships: (state includes county, state adjacency, county adjacency). In our experiments, the constructed THN spans from May 1 to 21, 2020, with 21 time snapshots, limiting each relationship type to a maximum of 2000 edges per snapshot.
\end{itemize}

\begin{table}[htbp]
	\centering
	\scriptsize
	\caption{Statistics of Datasets.}
	\resizebox{\linewidth}{!}{
		\begin{tabular}{ccccc}
			\toprule
			Dataset             & Math-overflow & Taobao & OGBN-MAG & COVID-19 \\
			\midrule
			\# Nodes            & 24818         & 29475  & 17269    & 2165     \\
			\# Edges            & 506550        & 63367  & 40000    & 131649   \\
			\# Snapshots        & 11            & 5      & 10       & 21       \\
			Node Type           & 1             & 3      & 4        & 2        \\
			Edge Type           & 3             & 3      & 4        & 3        \\			
			Time Granularity    & Second        & Second & Year     & Day      \\
			\bottomrule
		\end{tabular}
	}
	\label{tab:datasets}%
\end{table}

\subsubsection{Dataset Processing}

After the time window is partitioned, further data cleansing is necessitated. In our model and the baselines, representation vectors of nodes are derived from the initial $T$ network snapshots.  However, it is not feasible to obtain representation vectors for nodes that fail to appear in the $(T+1)$-th network snapshots, rendering link prediction for these nodes impossible. Consequently, nodes that newly appear, along with their corresponding links in the $(T+1)$-th snapshot, are eliminated. 
Additionally, the negative links for the training set are collected from links  that are absent in the initial $T$ snapshots.  For the test set, negative links are sampled from those not present in the $(T+1)$-th snapshot.
The node representations are derived from the first $T$ network snapshots $\{\mathcal{G}^1, \mathcal{G}^2, \cdots ,\mathcal{G}^T\}$. The links in the subsequent network snapshot $\mathcal{G}^{T+1}$ serve as the evaluation set. Moreover, 20\% of the links in this evaluation set are randomly assigned as the validation set, another 20\% as the positive training set, and the remaining 60\% as the positive test set. Simultaneously, several non-existent negative links are randomly sampled to form negative training and test sets.

\subsubsection{Metrics}
We employ the widely-adopted metrics AUC and AP for evaluation. AUC represents the area under the Receiver Operating Characteristic (ROC) curve, which plots the false positive rate on the x-axis and the true positive rate on the y-axis. AP refers to the area under the Precision-Recall curve, with recall on the  horizontal axis and precision on the vertical axis. Values of AUC and AP closer to 1 indicate superior model performance.

\subsection{Baselines}
\label{sec:baseline_appendix}

The baselines encompass both traditional and advanced link prediction models that are highly relevant to our research.
Detailed descriptions of the fundamental elements of the baseline models and our \mymodel are presented in Table \ref{tab:baselineModel}.

\textbf{(1) Static Homogeneous approaches}:
\begin{itemize}[leftmargin=*]
	\item \textbf{SEAL}~\cite{zhang2018link} extracts local graphs for the target link and learns their features to estimate the probability of that link's existence.
	
	\item \textbf{VGNAE}~\cite{ahn2021variational} integrates variational inference, GCNs, and normalization techniques to effectively learn probabilistic node embeddings from graph-structured data. 
\end{itemize}

\textbf{(2) Static Heterogeneous approaches}: 
\begin{itemize}[leftmargin=*]
	\item \textbf{Metapath2Vec}~\cite{dong2017metapath2vec} leverages metapath-guided random walks combined with the skip-gram model to learn low-dimensional embeddings for nodes in heterogeneous information networks, capturing both structural and semantic relationships within the network.
	
	\item  \textbf{GATNE}~\cite{cen2019representation} learns node representations in attributed multiplex networks. It successfully addresses the challenges of integrating multiple types of relationships and node attributes into a unified representation.
\end{itemize}

\textbf{(3) Dynamic Homogeneous approaches}: 
\begin{itemize}[leftmargin=*]
	\item \textbf{TGAT}~\cite{XuRKKA20} presents an inductive representation learning approach for temporal graphs, combining temporal encoding and GNNs to capture the dynamic nature of such graphs, which can be generalized to new nodes and future graph snapshots.
	\item \textbf{TDGNN}~\cite{qu2020continuous} designs a continuous-time link prediction method for dynamic graphs, leveraging temporal encodings and attention mechanisms to enhance the predictive capabilities of GNNs.
\end{itemize}

\textbf{(4) Dynamic Heterogeneous approaches}:
\begin{itemize}[leftmargin=*]
	\item \textbf{THAN}~\cite{li2023memory} leverages memory mechanisms and transformer architectures to capture intricate temporal and structural information of temporal heterogeneous graphs.
	
	\item \textbf{THGAT}~\cite{zhang2023dynamic} combines neighborhood type modeling, neighborhood information aggregation, and time encoding technique to achieve accurate node representations.
\end{itemize}

\textbf{(5) Ours}:
\begin{itemize}[leftmargin=*]
	\item \mymodel \textit{without Node-level } \textbf{(\clp$^{-N}$)}, which excludes the the node-level heterogeneity differentiation loss to verify its enhancement for the node-level node embedding. 
	
	\item \mymodel \textit{without Edge-level} \textbf{(\clp$^{-E}$)}, which removes the the edge-level heterogeneity differentiation loss to validate its efficacy in enhancing edge-level node embedding.
	
	\item \mymodel \textit{without Time-level} \textbf{(\clp$^{-T}$)}, which eliminates the the time-level heterogeneity differentiation loss to examine its contributions to temporal information modeling.
\end{itemize}

\end{document}